\documentclass[12pt]{article}

\usepackage[margin=1in]{geometry}
\usepackage{setspace}
\usepackage{amsmath,amssymb}
\usepackage{graphicx}
\usepackage{xcolor}

\usepackage{booktabs}
\usepackage{array}
\usepackage[normalem]{ulem}
\usepackage{longtable}
\usepackage{tikz}
\usepackage{enumitem}
\usepackage{xurl}
\usepackage{verbatim}
\usepackage{caption}
\usepackage{tabularx}
\usepackage{listings}

\usepackage[version=4]{mhchem}
\usepackage{float}
\usepackage{placeins}
\usepackage[numbers,sort&compress,super]{natbib}

\usepackage[colorlinks=true, urlcolor=blue, linkcolor=blue, citecolor=blue]{hyperref}

\definecolor{vscodebg}{RGB}{255, 255, 255}
\definecolor{vstext}{RGB}{30, 30, 30}
\definecolor{vscomment}{RGB}{0, 128, 0}
\definecolor{vsstring}{RGB}{163, 21, 21}
\definecolor{vskeyword}{RGB}{0, 0, 255}
\definecolor{vslinenum}{RGB}{128, 128, 128}
\definecolor{vsborder}{RGB}{200, 200, 200}

\lstdefinelanguage{json}{
  keywords={true,false,null},
  keywordstyle=\color{vskeyword}\bfseries,
  commentstyle=\color{vscomment}\itshape,
  stringstyle=\color{vsstring},
  morestring=[b]",
  basicstyle=\ttfamily\footnotesize\color{vstext},
  backgroundcolor=\color{vscodebg},
  frame=single,
  rulecolor=\color{vsborder},
  numbers=left,
  numberstyle=\tiny\color{vslinenum},
  breaklines=true,
  tabsize=2,
  showstringspaces=false
}

\lstdefinelanguage{yaml}{
  morekeywords={name,version,description},
  keywordstyle=\color{vskeyword},
  commentstyle=\color{vscomment}\itshape,
  stringstyle=\color{vsstring},
  morestring=[b]",
  moredelim=[l][\color{vstext}]{-},
  moredelim=[l][\color{vstext}]{:},
  basicstyle=\ttfamily\footnotesize\color{vstext},
  backgroundcolor=\color{vscodebg},
  frame=single,
  rulecolor=\color{vsborder},
  numbers=left,
  numberstyle=\tiny\color{vslinenum},
  breaklines=true,
  tabsize=2,
  showstringspaces=false
}

\begin{document}


\title{UniMatSim: An automated framework for high-throughput materials simulations using universal machine learning interatomic potentials}

\author{%
\begin{minipage}{0.96\textwidth}
\centering
Yanjin Xiang$^{1}$, Yihan Nie$^{1}$, Weiyi Wang$^{3}$ and Yunzhi Gao$^{1}$\\[0.25em]
Haidi Wang$^{2,*}$ and Wei Hu$^{3,*}$\\[0.65em]
{\small
$^{1}$School of Emerging Technology, University of Science and Technology of China, Hefei, Anhui 230026, China\\
$^{2}$School of Physics, Hefei University of Technology, Hefei, Anhui 230009, China\\
$^{3}$State Key Laboratory of Precision and Intelligent Chemistry, University of Science and Technology of China, Hefei, Anhui 230026, China\\[0.35em]
$^{*}$Correspondence: Haidi Wang (\texttt{haidi@hfut.edu.cn}); Wei Hu (\texttt{whuustc@ustc.edu.cn})
}
\end{minipage}}
\date{}
\maketitle

\begin{abstract}
Universal machine learning interatomic potentials (UMLIPs) reduce the cost of atomistic simulations, but heterogeneous interfaces and workflow implementations complicate high-throughput use. We present UniMatSim, a Python framework that unifies access to multiple UMLIPs and provides reusable modules for structural optimization, elastic properties, phonon calculations and molecular dynamics. Its configuration-driven engine represents task dependencies as directed acyclic graphs, executes independent tasks in parallel, and supports checkpointing and resumption. UniMatSim also provides command-line tools, remote model invocation, data conversion and utilities for two-dimensional materials. We demonstrate its use by screening two-dimensional Lieb-lattice derivatives. Four independently initialized Uni2D instances screen 1,176 structures and retain 393 consensus candidates under the adopted structural, elastic and phonon criteria. Subsequent density functional theory calculations and spin-resolved band analysis identify 59 selected candidates with spin-resolved band features consistent with altermagnetic behavior. UniMatSim provides a reusable framework for high-throughput UMLIP-based materials simulations.
\end{abstract}

\section{Introduction}

Materials science is undergoing a paradigm shift driven by data-driven methods\cite{himanen2019datadriven,sivan2024advances}. Materials discovery has long relied heavily on empirical knowledge and repeated trial-and-error experiments. This approach is time-consuming and costly, and it makes systematic exploration of the vast design space defined by composition, structure, and processing conditions difficult. Advances in computing power and theoretical methods have made high-throughput computational simulations an important tool for addressing these limitations\cite{daglar2020recent,polat2020co2,fowler2024beyond}.

Through standardized, automated workflows, high-throughput calculations can rapidly predict and evaluate the physical, chemical, and functional properties of thousands of candidate materials, enabling efficient screening and prioritization. This computationally driven, experimentally guided research paradigm\cite{jakob2026learning} substantially shortens materials development cycles, reduces costs, and decreases uncertainty during early-stage exploration.
Its impact is particularly evident in key areas such as gas adsorption and separation in porous materials\cite{palakkal2024exploiting}, semiconductors, industrial catalysts, and structural alloys, including high-entropy alloys\cite{jain2024leveraging,ozdemir2024machine}. These developments open new possibilities for rational, performance-oriented materials design and have considerable scientific and strategic significance.

In recent years, a new generation of machine learning techniques, particularly universal machine learning interatomic potentials (UMLIPs), has advanced rapidly in materials simulations, reshaping the traditional research paradigm centered on first-principles calculations\cite{batzner2022E3}. Models such as CHGNet\cite{deng2023chgnet}, M3GNet\cite{chen2022universal}, MACE\cite{batatia2022mace}, MatterSim\cite{yang2024mattersim}, and SevenNet\cite{park2024scalable} can substantially reduce computational costs while retaining good agreement with DFT energies and atomic forces within their respective training and benchmark domains\cite{chen2022universal,deng2023chgnet}. They offer new approaches to large-scale materials screening and dynamical simulations across length scales and chemical compositions.

Workflow infrastructures such as atomate2\cite{rosen2024atomate2},
jobflow\cite{rosen2024jobflow}, and AiiDA\cite{huber2022automated}
already provide mature automation and provenance capabilities and support
machine learning force fields and first-principles calculations to varying
degrees. However, applying these advanced potentials efficiently, reliably,
and at scale in practical materials research still presents several
challenges. First, the software ecosystem remains fragmented: models are
independently developed and maintained by different teams without a common
software architecture or development conventions. Second, differences in
computational procedures, model loading, parameter configuration, and
fine-tuning and inference interfaces increase the learning effort and
development costs for researchers. The absence of a shared abstraction
layer and standardized interfaces also complicates systematic model
comparison, flexible switching, and joint integration, limiting broader
use of these potentials.

To address these challenges, we present UniMatSim, an integrated Python
framework for high-throughput materials simulations using UMLIPs.
At the model layer, a unified PotentialModel interface encapsulates
multiple potential models, including CHGNet\cite{deng2023chgnet},
M3GNet\cite{chen2022universal}, MACE\cite{batatia2022mace},
MatterSim\cite{yang2024mattersim}, and SevenNet\cite{park2024scalable},
among others. Users can switch models through simple configuration changes without rewriting scripts or adapting to different calling conventions.
At the task layer, common computational steps, including structural optimization, elastic constant calculations, phonon analysis, and molecular dynamics simulations, are represented as standardized, reusable Task modules with clearly defined inputs, outputs, and execution logic. At the workflow layer, WorkflowEngine automates task orchestration, dependency resolution, parallel scheduling, and state management, supporting workflows ranging from simple linear sequences to complex directed acyclic graphs (DAGs). At the computational engine layer, UniMatSim wraps the ASE Calculator interface to provide consistent access to different potential models and offers REST-based remote computational services through which clients can access server-side models in a consistent manner. This design aims to improve automation, reproducibility, and overall computational efficiency in materials simulations, allowing researchers to devote more effort to scientific questions and less to workflow management and implementation.

The following sections describe UniMatSim's configuration-driven execution architecture and key implementations, including the unified potential model interface, task-graph scheduling, command-line interface (CLI), remote model invocation, and handling of two-dimensional materials. A representative application to high-throughput stability screening of two-dimensional Lieb lattices then illustrates the complete workflow, computational efficiency, and robustness in practical use. Finally, we discuss the framework's scope and future development.

\section{Results}

\subsection{Configuration-driven execution architecture}

\begin{figure}[!htbp]
\centering
\includegraphics[width=1.00\textwidth]{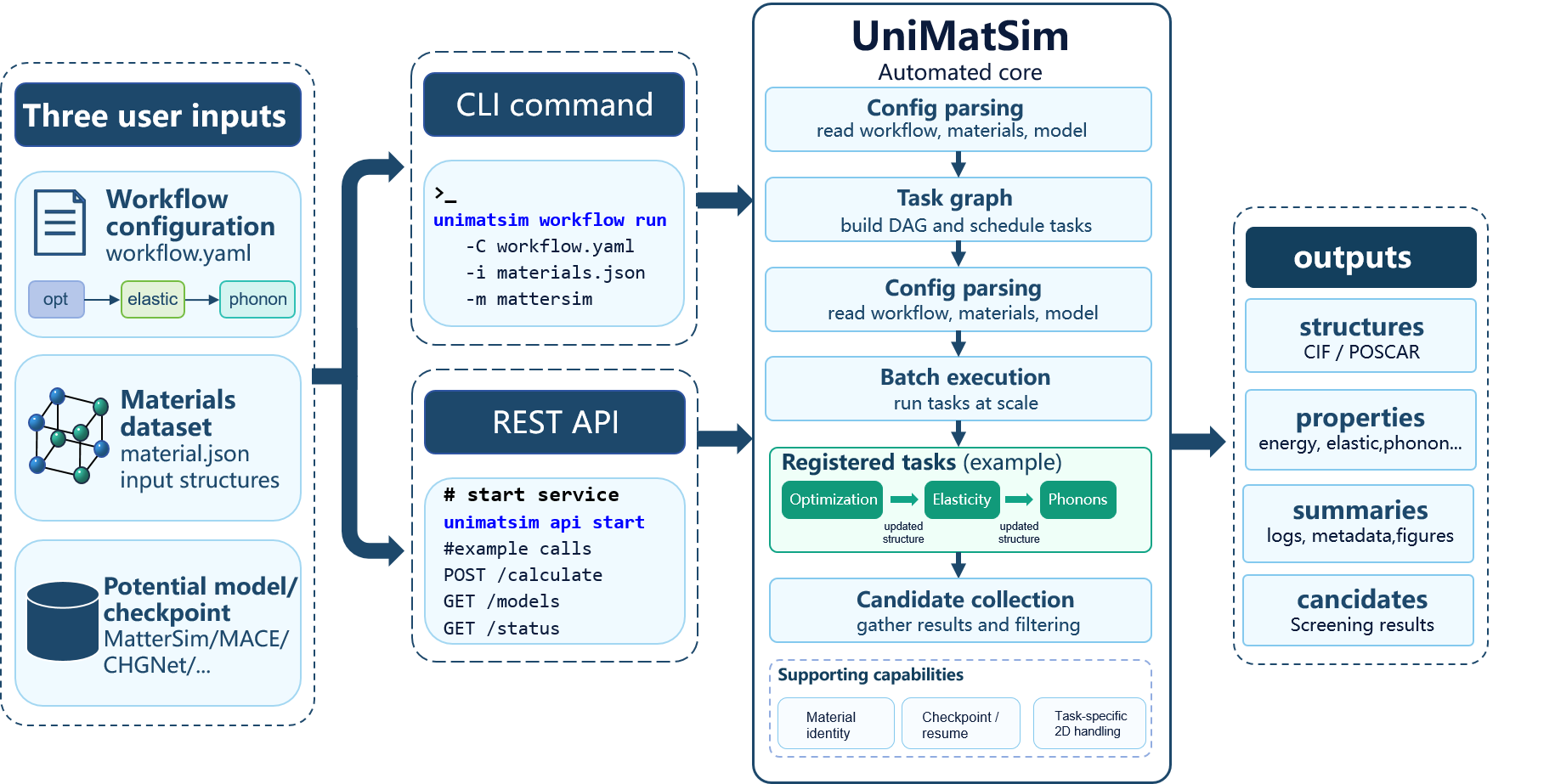}
\caption{\textbf{Configuration-driven execution architecture of UniMatSim.} Users provide a workflow configuration, a material dataset, and a potential model or checkpoint through the CLI. The framework constructs a task dependency graph, initializes the selected computational backend, executes registered tasks, propagates updated structures between dependent tasks, and aggregates material-level results. Checkpointing supports workflow resumption, while specialized modules handle two-dimensional materials. An ASE-compatible \texttt{RemoteCalculator} provides access to server-side potential models through the REST API.}
\label{fig:architecture}
\end{figure}

UniMatSim's user-facing execution model starts from three inputs: a workflow configuration, a material dataset, and a potential model or model checkpoint. For multistage screening, users bind these inputs to a single run through \lstinline{unimatsim workflow run}. The framework then parses the workflow definition, constructs a task dependency graph, initializes the selected computational backend, executes registered tasks over the material dataset, and passes updated structures from upstream tasks to dependent tasks. Material identifiers, task states, and checkpoint information are retained during execution. Results are ultimately aggregated into structure files, property data, run summaries, and screened candidates. Figure~\ref{fig:architecture} illustrates this configuration-driven execution path and the entry point for remote potential model invocation.

The automated execution in Figure~\ref{fig:architecture} proceeds through several successive steps. The configuration parser first reads the workflow, materials, and model information, and \lstinline{WorkflowBuilder} converts tasks and their dependencies into a DAG. The computational engine then initializes a calculator using the user-specified potential model or checkpoint. Registered tasks execute in dependency order at the workflow layer. Updated structures produced by structural optimization can be passed directly to downstream elastic and phonon tasks, while other property results, file paths, and execution states are retained in the workflow context. After batch execution, the results layer aggregates material-level outputs and generates candidate sets according to task-specific screening criteria.

The implementation comprises modules with distinct responsibilities: \lstinline{core} provides shared definitions, configuration, and error handling; \lstinline{structures} handles structures and low-dimensional geometry; \lstinline{engine} encapsulates potential models and calculator calls; \lstinline{workflows} manages task graphs, dependency propagation, and runtime state; \lstinline{data} handles external DFT data and training-data conversion; and \lstinline{interface} and the CLI provide programmatic and command-line entry points.

The CLI/workflow interface is the primary entry point for local multistage screening. Remote potential evaluation is also available through a REST API and an ASE-compatible RemoteCalculator, as described below.

\subsection{Key technical implementations}

This section describes the key implementations underlying the execution chain in Figure~\ref{fig:architecture}. We first introduce the unified potential model interface and registered task abstraction, which are decoupled from specific UMLIP backends. We then describe task-graph construction, structure and state propagation, batch execution, and the CLI workflow entry point, followed by REST-based remote model invocation and the data extraction and fine-tuning tools used for model preparation.

\subsubsection{Unified machine learning interatomic potential interface}

A key feature of UniMatSim is its unified interface to different interatomic potentials, with consistent potential management implemented through the CalculationInvocation class. UniMatSim uses the Atomic Simulation Environment (ASE)\cite{ASE} as its primary calculator backend. For example, when the Lennard-Jones potential\cite{lennardjones1924} is selected, the engine module instantiates ASE's LennardJones calculator and associates it with the atomic structure. For machine learning potentials such as M3GNet and CHGNet, UniMatSim can call their native Python implementations directly or wrap them through ASE's external interfaces where available. All calculators follow a common interface for evaluating energies, forces, stresses, and related properties. This abstraction decouples higher-level workflow tasks from model-specific implementations, preserving model independence. The design follows ASE's unified Calculator concept and extends it within a complete workflow framework. Users can therefore switch between models by calling set\_potential\_model(), without modifying downstream calculation code. The interface currently supports:

\begin{itemize}
\item UMLIPs: CHGNet, M3GNet, MACE, MatterSim, SevenNet, DeepMD\cite{deepmd}
\item Classical potentials: Lennard-Jones and Buckingham\cite{buckingham1938classical} (for testing and validation)
\end{itemize}

\subsubsection{Modular task architecture and computational capabilities}
To support extensibility and maintainability, UniMatSim adopts a modular architecture that represents complex simulation procedures as discrete, independent units called \textit{Tasks}. Simulation logic is dynamically registered using decorators rather than hard-coded into the core. This design decouples the definition of physical calculations from the workflow execution engine, allowing new computational methods to be integrated without modifying the core infrastructure.

A \textit{Task} in UniMatSim can orchestrate an entire multistage physical workflow rather than a single elementary operation. For example, an elastic constant calculation (ELASTIC) is an automated procedure comprising structural relaxation, symmetry-adapted strain generation, energy and force evaluation, and stress--strain fitting. This higher-level abstraction supports methodological consistency and reproducibility across high-throughput screening tasks and reduces the likelihood of manual errors in complex simulation chains.

UniMatSim currently integrates nine core task modules, grouped into four functional categories:

\textbf{A. Structural optimization and symmetry analysis}
\begin{enumerate}[label=\textbf{\arabic*.}]
\item \textbf{Geometry optimization (OPTIMIZATION):} Provides optimizers including BFGS, FIRE\cite{bitzek2006structural}, and LBFGS (see\cite{ASE} for their ASE implementations), with configurable force thresholds and maximum optimization steps. 
\item \textbf{Symmetry characterization (SYMMETRY):} Integrates \lstinline{spglib} for automatic space-group identification and primitive-cell standardization to reduce computational costs.\cite{spglib}
\item \textbf{Single-point calculations (SINGLE\_POINT):}
Computes energies, atomic forces, and stress tensors to support subsequent workflow tasks.
\end{enumerate}

\textbf{B. Mechanical and dynamical properties}
\begin{enumerate}[label=\textbf{\arabic*.}]
\item \textbf{Elastic stiffness tensor (ELASTIC):} Automatically calculates second-order elastic constants using the stress--strain method, derives bulk, shear, and Young's moduli, and rigorously checks mechanical stability criteria.\cite{nielsen1985quantum, lepage2002symmetry, born1954dynamical, mouhat2014necessary, wei2014superior}
\item \textbf{Equation of state (EOS):} Fits energy--volume data to the Birch--Murnaghan equation of state to determine equilibrium lattice parameters and the bulk modulus under hydrostatic conditions.\cite{birch1947finite}
\item \textbf{Phonon dispersion (PHONON):} Uses the finite-displacement method\cite{togo2015first, parlinski1997first} to calculate phonon spectra and the density of states (DOS), with automatic generation of high-symmetry paths for Brillouin-zone sampling\cite{hinuma2017band}.
\item \textbf{Molecular dynamics (MOLECULAR\_DYNAMICS):} Supports finite-temperature simulations in the NVE ensemble and mean-squared displacement (MSD) analysis of thermodynamic evolution and transport phenomena.
\end{enumerate}

\textbf{C. Thermal transport properties}
\begin{enumerate}[label=\textbf{\arabic*.}]
\item \textbf{Boltzmann transport solver (BTE):} Interfaces with \lstinline{phono3py} to calculate lattice thermal conductivity by solving the linearized Boltzmann transport equation and accounting for anharmonic phonon--phonon interactions.\cite{phono3py, togo2015distributions, chaput2013direct, ziman2001electrons}
\end{enumerate}

\textbf{D. Data extraction}
\begin{enumerate}[label=\textbf{\arabic*.}]
\item \textbf{Data import (VASP\_EXTRACT):}
Automatically parses batches of existing VASP outputs and standardizes the extracted structures and physical quantities. The resulting data can be used to prepare datasets for fine-tuning machine learning potentials or to provide input structures for high-throughput calculations.
\end{enumerate}

\subsubsection{Workflow scheduling strategy (WorkflowBuilder)}

A multistage screening run is defined by an input structure set $\mathcal{S}$, a selected set of potential models or model variants $\mathcal{M}$, and a task graph $\mathcal{G}=(\mathcal{V},\mathcal{E})$. The task graph is a DAG in which each node $v\in\mathcal{V}$ represents a task with explicit inputs, calculation parameters, and outputs, and each edge $(v_i,v_j)\in\mathcal{E}$ indicates that task $v_j$ depends on task $v_i$. Users declare tasks and dependencies in the workflow configuration; \lstinline{WorkflowBuilder} constructs the graph, and \lstinline{WorkflowManager} organizes execution in dependency order. For tasks with structural dependencies, updated upstream structures are passed directly downstream. Other property results, file paths, and execution states are retained in the workflow context for dependency checks, candidate aggregation, and result provenance. Thus, tasks, parameters, and execution order in a screening protocol can be defined independently of model invocation and batch execution code and reused across material datasets and supported models.

WorkflowBuilder constructs execution phases using a layered topological
sort based on Kahn's algorithm. Initially, all tasks with zero in-degree
are added to a ready queue. At each iteration, all tasks currently in
the queue form one execution phase. The algorithm then decrements the
in-degrees of their successors, adding a successor to the next phase
only after all its predecessors have been processed. This procedure
ensures that the resulting phases respect all task dependencies.

At runtime, WorkflowManager executes these phases sequentially.
When \lstinline{executor="parallel"} is selected, tasks within the same
phase are executed concurrently through a thread pool; otherwise,
they are executed serially. Execution advances to the next phase only
after all tasks in the current phase have completed successfully,
introducing a synchronization barrier between phases. The ordering
of independent tasks within a phase is not guaranteed.

For example, a workflow in which structural relaxation precedes
independent elastic and phonon calculations, followed by a stability
assessment that depends on both, is organized into three phases:
relaxation, elastic and phonon calculations, and stability assessment.
With parallel execution enabled, the elastic and phonon tasks can run
concurrently, while the stability assessment starts after their shared
execution phase has completed successfully.

The framework also implements checkpointing, periodically persisting intermediate results to disk or a database and associating them with a unique workflow hash ID. If a task fails or a run is interrupted, execution can resume from the most recent valid checkpoint without restarting the entire workflow.

\subsubsection{Command-line interface and interaction design}

The CLI is the primary user entry point for the configuration-driven workflow in Figure~\ref{fig:architecture}. UniMatSim's command-line interface is built on Typer\cite{typer} and organizes functionality into property calculations (\texttt{calc}), workflow orchestration (\texttt{workflow}), data import (\texttt{data}), and post-processing analysis (\texttt{analysis}). For multistage screening, \texttt{workflow run} binds the workflow configuration, material inputs, and potential model to a single run. For individual property evaluations, \texttt{calc} provides direct task entry points. This design allows the three primary user inputs in Figure~\ref{fig:architecture} to enter a common execution framework through a unified command.

\paragraph{Direct property calculation module}

The \texttt{calc} module organizes the execution of basic physical property solvers and provides a primary entry point for atomistic simulations, supporting both individual tasks and parallel batch execution. Its main capabilities include:

\begin{enumerate}[label=\textbf{\arabic*.}]
\item \textbf{Geometry optimization:} The system can rapidly determine ground-state structures. For example, \lstinline{unimatsim calc optimize -i Si.cif -m chgnet} relaxes an individual structure. Batch processing can be enabled with \lstinline{unimatsim calc optimize -i ./structures/ -m mattersim --parallel 4}, distributing optimization tasks across four concurrent processes to utilize multicore nodes.

\item \textbf{Elastic stiffness tensor:} The command \lstinline{unimatsim calc elastic -i optimized.cif -m mattersim} automatically applies a deformation--energy method to obtain the elastic tensor. The \lstinline{--check-stability} flag evaluates Born--Huang mechanical stability criteria during the calculation and filters out unstable candidates immediately afterward.

\item \textbf{Phonon dispersion:} Vibrational properties are calculated using the finite-displacement method. In \lstinline{unimatsim calc phonon} \lstinline{-i structure.cif} \lstinline{--supercell 2,2,2} \lstinline{--mesh 10,10,10} \lstinline{-m chgnet}, \lstinline{--supercell} specifies the replication factors along the three lattice vectors, and \lstinline{--mesh} sets the phonon density-of-states $q$-point mesh. Force constants and phonon band structures are generated automatically.

\item \textbf{Molecular dynamics:} Commands such as \lstinline{unimatsim calc md -i initial.cif --ensemble nvt --temperature 300 --steps 10000 -m m3gnet} simulate time evolution at finite temperature. The module supports common ensembles (NVE, NVT, and NPT) and detailed control of thermodynamic parameters for studies of phase stability and diffusion dynamics.
\end{enumerate}

\paragraph{Automated workflow orchestration module}

The \texttt{workflow} module manages complex simulations by representing tasks as nodes in a DAG and uses declarative configuration files (YAML/JSON) for dependency resolution and concurrent execution.

\begin{enumerate}[label=\textbf{\Alph*.}]
\item \textbf{Declarative configuration generation:} Generator commands allow users to construct complex workflows rapidly. For example, \lstinline{unimatsim workflow create} \lstinline{-o workflow.yaml} \lstinline{-s opt,elastic,phonon} \lstinline{-m mattersim} generates a complete configuration for a sequential relaxation--elastic--phonon workflow. The resulting YAML file shown below contains all execution parameters and dependency logic:

\begin{lstlisting}[language=yaml, breaklines=true, breakindent=0pt]
metadata:
  description: "Sequential workflow generated from: opt, elastic, phonon"
  ml_potential: "mattersim"
  ml_params:
    model_path: "best_model.pth"
  potential_type: "ml"

tasks:
  - name: "optimization"
    type: "optimize"
    config:
      fmax: 0.01
      steps: 500
      optimizer: "BFGS"

  - name: "elastic"
    type: "elastic"
    config:
      fmax: 0.1
      relax_structure: true
      check_stability: true
      axial_strain_set: [-0.01, -0.005, 0.005, 0.01]
      shear_strain_set: [-0.06, -0.03, 0.03, 0.06]

  - name: "phonon"
    type: "phonon"
    config:
      supercell: [2, 2, 2]
      dos_kpts: [10, 10, 10]
      check_stability: true
      save_bandstructure: true

dependencies:
  optimization: "elastic"
  elastic: "phonon"
  phonon: null
\end{lstlisting}

\item \textbf{Workflow execution and scheduling:} The command \lstinline{unimatsim workflow run -C workflow.yaml -i ./structures/} launches the workflow engine. The system topologically sorts the task graph and executes independent nodes in parallel where possible. State persistence is handled automatically, and a \lstinline{.checkpoint_<workflow_id>.json} log records task completion.

\item \textbf{Fault tolerance and recovery:} To reduce the impact of interruptions during long simulations, the framework supports state recovery through \lstinline{unimatsim workflow resume} \lstinline{--id <workflow_id>} \lstinline{-C workflow.json}. Execution resumes from the most recent successful checkpoint, substantially improving efficiency in unstable computing environments.
\end{enumerate}

\paragraph{Data extraction and conversion module}

The \texttt{data} module implements an extract, transform, and load (ETL) process that connects raw first-principles data with machine learning training requirements.

\begin{enumerate}[label=\textbf{\arabic*.}]
\item \textbf{High-throughput extraction:} The command \lstinline{unimatsim data import vasp -i ./vasp_results/ -o extracted_data.json} parses directories containing large numbers of VASP outputs and extracts key physical quantities, including energies, atomic forces, and stress tensors. The \lstinline{--converged-only} filter discards unconverged results to preserve data quality.

\item \textbf{Structure serialization:} For batch structure processing, \lstinline{unimatsim data import poscar -i ./POSCAR_files/ -o structures.csv --format csv} aggregates lattice parameters and atomic coordinates from multiple POSCAR files and converts them into a tabular format suitable for statistical analysis.

\item \textbf{Training-data standardization:} To facilitate model training, \lstinline{unimatsim data convert to-xyz} \lstinline{-i ./structures/} \lstinline{-O ./training_data/} converts structure datasets into Extended XYZ, a standard input format for many modern machine learning potential frameworks.

\item \textbf{Dataset partitioning:} Large datasets can be managed using \lstinline{unimatsim data slide -i large_dataset.json -o chunks/ --chunk-size 1000}. This command partitions data by record count into manageable chunks, optimizing memory use during subsequent training or analysis.
\end{enumerate}

\paragraph{Post-processing and validation module}

The \texttt{analysis} module provides tools for structural characterization and comparative validation of model predictions.

\begin{enumerate}[label=\textbf{\arabic*.}]
\item \textbf{Symmetry identification:} The command \lstinline{unimatsim analysis symmetry} \lstinline{-i optimized_structures/} \lstinline{--symprec 0.01} uses \lstinline{spglib} to identify crystal systems, space groups, and point groups. The \lstinline{--symprec} parameter controls the numerical tolerance, while \lstinline{--filter} enables selective structure screening based on symmetry constraints.

\item \textbf{Consensus stability analysis:} The command \lstinline{unimatsim analysis compare module -i ./results/ -o stability_comparison.png} compares screening results from multiple potential models or model variants and generates an UpSet plot\cite{lex2014upset} showing overlaps between candidate sets. Each model independently performs prescreening using identical input structures, task graphs, and calculation parameters. For model $m\in\mathcal{M}$, let $\mathcal{C}_m$ denote the candidates that pass screening. Their strict intersection is
\[
\mathcal{C}_{\mathrm{cons}}=\bigcap_{m\in\mathcal{M}}\mathcal{C}_m .
\]
This set retains structures that satisfy the same screening criteria under every model included in the comparison and can serve as a candidate pool for subsequent DFT validation.
\end{enumerate}
\subsubsection{REST API and remote potential model invocation}

In addition to local CLI/Python execution, UniMatSim provides remote potential model evaluation through a REST service built on FastAPI\cite{fastapi, narayanan2024engineering}. The service handles model invocation rather than complete workflow scheduling: clients submit atomic numbers, atomic coordinates, the unit cell, periodic boundary conditions, and the calculation type; the server invokes the specified potential model and returns quantities such as energies, atomic forces, and stresses. The server can select a model for each request or use a default model. Model caching reuses initialized backends to reduce repeated loading overhead across successive requests.

The client-side \lstinline{RemoteCalculator} is compatible with ASE's \lstinline{Calculator} interface. It serializes ASE \lstinline{Atoms} objects into REST requests and maps the energies, forces, and stresses returned by the server to calculator results. UniMatSim's internal \lstinline{ModelInvocation} distinguishes \lstinline{DIRECT} and \lstinline{REMOTE} invocation modes, enabling programs based on the unified calculator abstraction to access remote models without changing higher-level property calculation logic. The Remote client--REST API branch in Figure~\ref{fig:architecture} represents this remote model invocation path.

The REST service provides endpoints including \texttt{/calculate}, \texttt{/models}, and \texttt{/status}, with optional API-key/JWT authentication and request rate limiting. The service can be started using the following command, after which remote clients or UniMatSim's API entry point can submit calculation requests:
\begin{lstlisting}[language=bash, breaklines=true, breakindent=0pt]
unimatsim api start --host 0.0.0.0 --port 8000
unimatsim api call calculate -i structure.cif --url http://remote-host:8000 --type all
\end{lstlisting}

The REST service evaluates individual potential-model requests, while the local workflow layer manages task dependencies, batch processing, checkpointing, resumption, and candidate aggregation. Cluster-level resource management is handled by the external high-performance computing (HPC) environment.

\subsubsection{Training-data preparation and potential model fine-tuning}

Beyond the screening execution chain in Figure~\ref{fig:architecture}, UniMatSim provides data extraction and fine-tuning tools for model preparation. When users need to construct or update model checkpoints, the framework can convert first-principles outputs into training data and invoke the corresponding backend's fine-tuning interface through a unified CLI.

The \lstinline{unimatsim.data} module provides efficient data extraction and conversion tools for batch extraction of total energies, atomic forces, and stresses from VASP output files (OUTCAR and vasprun.xml). To preserve training-data quality, \lstinline{VASPDataExtraction} uses an energy-tolerance-based structure selection mechanism to remove redundant configurations from molecular dynamics trajectories or structural relaxation paths automatically. \lstinline{VASPDataConverter} then converts the extracted data into standard machine learning formats, such as JSON or Extended XYZ, and filters anomalous energies or excessive atomic forces using predefined thresholds.

After data preparation, a unified fine-tuning CLI (\lstinline{unimatsim model fine-tuning}) supports multiple universal potentials, including CHGNet and MatterSim. The interface provides validation support: users can specify an independent validation data file through \lstinline{--valid-material}, or automatically reserve a fraction of the training data as a validation set using \lstinline{--valid-split} (default: 0.1). The \lstinline{--early-stop-patience} parameter (default: 10) configures early stopping, terminating training when the validation loss fails to improve over a specified number of consecutive epochs to reduce overfitting. For finer control, users can supply a JSON configuration file through \lstinline{--config} to set advanced hyperparameters, including force and stress loss weights (\lstinline{force_loss_ratio}, \lstinline{stress_loss_ratio}), energy renormalization (\lstinline{re_normalize}), and the learning-rate decay step size (\lstinline{step_size}).

After fine-tuning, the system creates an output directory named \lstinline{ft_<model>_<name>}, containing the final fine-tuned model (\lstinline{<model>_finetuned.pth}), the best checkpoint (\lstinline{best_model.pth}), and periodically saved intermediate checkpoints. The resulting model can be referenced directly in subsequent workflow configurations.

\subsubsection{Low-dimensional material identification and stability analysis}

Low-dimensional materials, particularly two-dimensional (2D) systems, have physical properties that differ substantially from those of three-dimensional materials and require specialized handling during structure identification, simulation setup, and result analysis. UniMatSim therefore integrates dedicated two-dimensional support into its structures/materials\_2d module. The framework uses pymatgen's\cite{pymatgen} dimensionality analysis tools to determine the dimensionality of input crystal structures automatically. The algorithm analyzes atomic connectivity matrices and periodic boundary conditions and uses graph-theoretical methods to determine the spatial connectivity of bonding networks. A structure is classified as two-dimensional when it contains a continuous in-plane covalent bonding network and a substantial van der Waals gap in the perpendicular direction.

After dimensionality identification, UniMatSim automatically adjusts subsequent simulation steps. During structural relaxation, for example, apply\_2d\_cell\_filter() constrains the cell using ASE's FrechetCellFilter\cite{ASE}, with a default mask=[1,1,0,0,0,1]. Only in-plane lattice degrees of freedom and atomic positions are relaxed, preserving a physically meaningful interlayer separation. For phonon and Boltzmann transport equation (BTE) calculations, supercells are automatically changed from (n,n,n) to (n,n,1), avoiding unnecessary computational overhead in the nonperiodic direction. The get\_filtered\_2d\_bz\_path() function also generates Brillouin-zone paths suitable for two-dimensional materials by excluding high-symmetry points with nonzero $k_z$ components, avoiding artifacts arising from spurious out-of-plane dispersion.

In the multistage screening used here, elastic and dynamical stability are the main criteria entering the workflow decision chain. Functions such as stability\_test\_3d() and stability\_test\_2d() assess elastic tensors using Born--Huang conditions, and two-dimensional tasks provide specialized criteria for common in-plane lattices. The phonon analysis module checks for soft modes or imaginary frequencies to assess dynamical stability, with task parameters specifying numerical tolerance thresholds at high-symmetry points and along general paths. The Python analysis layer also provides optional thermodynamic post-processing based on formation energies and energies above the convex hull (see\cite{bartel2018physical} for background on temperature-dependent Gibbs energies and convex-hull analysis), including comparisons with Materials Project reference data\cite{jain2013commentary}. 

\subsection{Application to two-dimensional Lieb lattices}

\subsubsection{High-throughput screening workflow for Lieb lattices}

To demonstrate UniMatSim's application to high-throughput materials discovery, we construct a material generation and stability prescreening procedure for two-dimensional Lieb lattices and their derivatives. The Lieb lattice is a classical model for strongly correlated electrons\cite{lieb1989two}, with a characteristic flat-band electronic structure arising from its three-site primitive cell\cite{goldman2011topological}. It provides an important platform for investigating magnetism, superfluidity\cite{julku2016geometric}, and topological phenomena\cite{weeks2010topological}.

\begin{figure}[!htbp]
\centering
\includegraphics[width=0.80\textwidth]{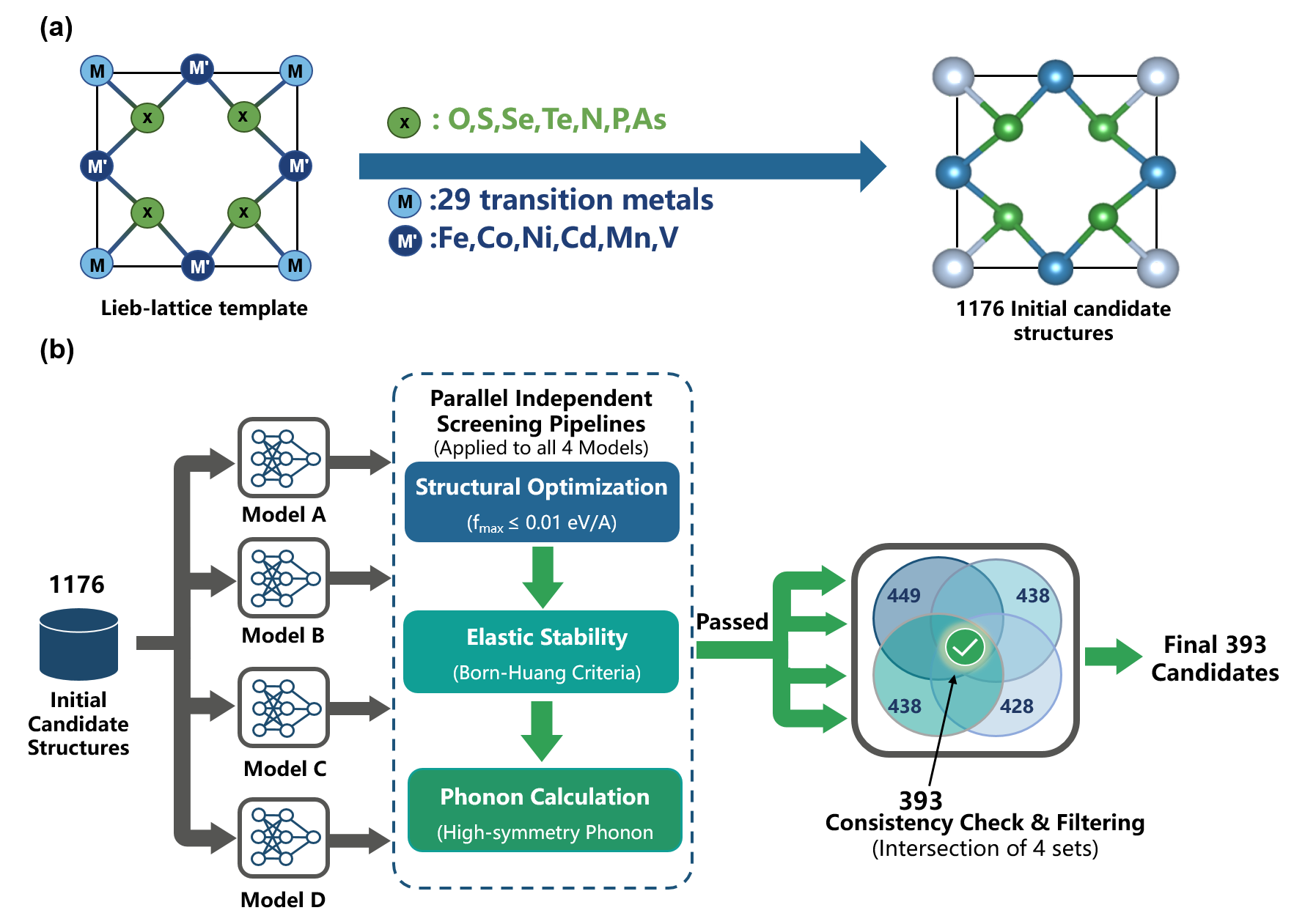}

\caption{\textbf{Lieb-lattice candidate generation and high-throughput UMLIP screening.}
(a) Elemental substitution of a two-dimensional Lieb-lattice template with the general formula $MM'_2X_4$. Light-blue, dark-blue, and green spheres represent the corner metal sites ($M$), edge-center metal sites ($M'$), and ligand sites ($X$), respectively. The $M$ sites are selected from 29 metals, $M'$ from Fe, Co, Ni, Cd, Mn, and V, and $X$ from O, S, Se, Te, N, P, and As. The representative TaCo$_2$Se$_4$ structure is shown in the upper-right panel, with Ta, Co, and Se represented by silver-white, blue, and green spheres, respectively.
(b) High-throughput screening using four Uni2D instances trained with different random seeds. Each instance performs structural optimization using ASE's \texttt{QuasiNewton} optimizer (\texttt{fmax = 0.01}, evaluated on cell-filtered generalized forces), elastic stability assessment based on Born--Huang criteria, and phonon screening along high-symmetry paths. Instances trained with seeds 42, 123, 2025, and 8888 retain 438, 438, 428, and 449 candidates, respectively, with 393 candidates in their strict intersection for subsequent DFT analysis. Colored regions schematically represent the individual candidate sets and their common intersection; overlap areas are not proportional to candidate counts.}

\label{fig:lieb_workflow}
\end{figure}

As shown in the upper part of Figure~\ref{fig:lieb_workflow}, the initial candidate set was generated by site-specific elemental substitution of a WFe$_2$Te$_4$ Lieb-lattice template, producing 1,176 structures with the general composition $MM'_2X_4$. The site-specific element pools and exclusion rule are detailed in Methods. The upper-right panel illustrates the substituted structure of TaCo$_2$Se$_4$, with Ta, Co, and Se occupying the $M$, $M'$, and $X$ sites, respectively.

UniMatSim then performs stability prescreening using Uni2D, a machine learning interatomic potential trained for two-dimensional materials\cite{2dmodel}. Four model instances trained with different random seeds independently execute the same structural optimization, elastic stability assessment, and phonon calculation procedure. Only structures passing one stage proceed to the next. The strict intersection of the four resulting sets contains 393 consensus candidates, which serve as inputs for subsequent independent DFT calculations of magnetic and electronic properties.

The workflow classifies candidates into three groups: those passing both elastic and dynamical screening, those classified as elastically unstable, and those passing elastic screening but failing dynamical screening. It generates structured reports containing elastic constants, phonon stability assessments, and related calculation results.

The same optimization--elastic--phonon workflow is independently executed with four Uni2D models trained using different random initializations, with training seeds 42, 123, 2025, and 8888. All runs use the calculation parameters and processing rules given above, with consistent material identifiers and output organization. They yield 438, 438, 428, and 449 candidate structures passing phonon screening, respectively.

To compare these candidate sets, UniMatSim's analysis module reads the candidate JSON files produced by each run, calculates two-, three-, and four-model intersections by material identifier, and generates the UpSet plot in Supplementary Information Figure~\ref{fig:upset_si}\cite{lex2014upset}. This procedure can be executed through \lstinline{unimatsim analysis compare module}, with intersection statistics also exportable in JSON format. Two-model intersections contain 404--421 candidates, three-model intersections contain 397--404 candidates, and the strict four-model intersection contains 393 candidates. These 393 structures form the consensus candidate pool for subsequent independent DFT calculations of magnetic and electronic properties.

\begin{figure}[!htbp]
  \centering
  \includegraphics[width=0.55\textwidth,height=0.58\textheight,keepaspectratio]{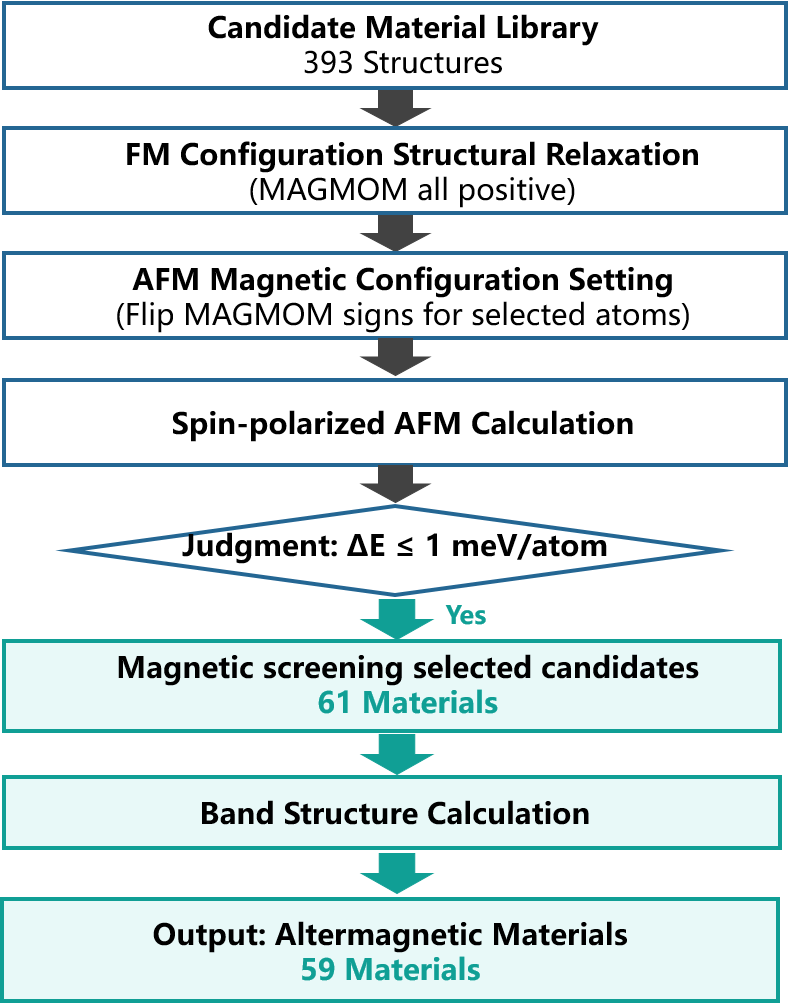}
  \caption{\textbf{DFT screening of magnetic and electronic properties.} Starting from the 393 candidates retained by the four-model intersection, reference geometries are relaxed using FM-like initial magnetic configurations. AFM configurations are constructed by reversing selected initial atomic magnetic moments and evaluated through spin-polarized single-point calculations on the reference geometries. Magnetic screening combines an energy criterion, $\Delta E \leq 1\ \mathrm{meV/atom}$, with inspection of the self-consistent local magnetic moments, where $\Delta E$ denotes the AFM energy relative to the FM-like reference per atom. Subsequent spin-resolved band calculations identify altermagnetic candidates.}
  \label{fig:lieb_DFT}
\end{figure}
Figure~\ref{fig:lieb_DFT} summarizes the DFT analysis of the magnetic and electronic properties of the consensus candidates. Magnetic screening combines energy comparisons between the tested collinear configurations with checks of the self-consistent local magnetic moments. Spin-resolved band calculations are then used to examine momentum-dependent spin splitting. The band structures of the 59 selected candidates are provided in Section~\ref{sec:si-band-structures} of the Supplementary Information, with computational details given in Section~\ref{sec:dft-calculations} of the main text.

\begin{figure}[!htbp]
  \centering
    \includegraphics[width=\linewidth]{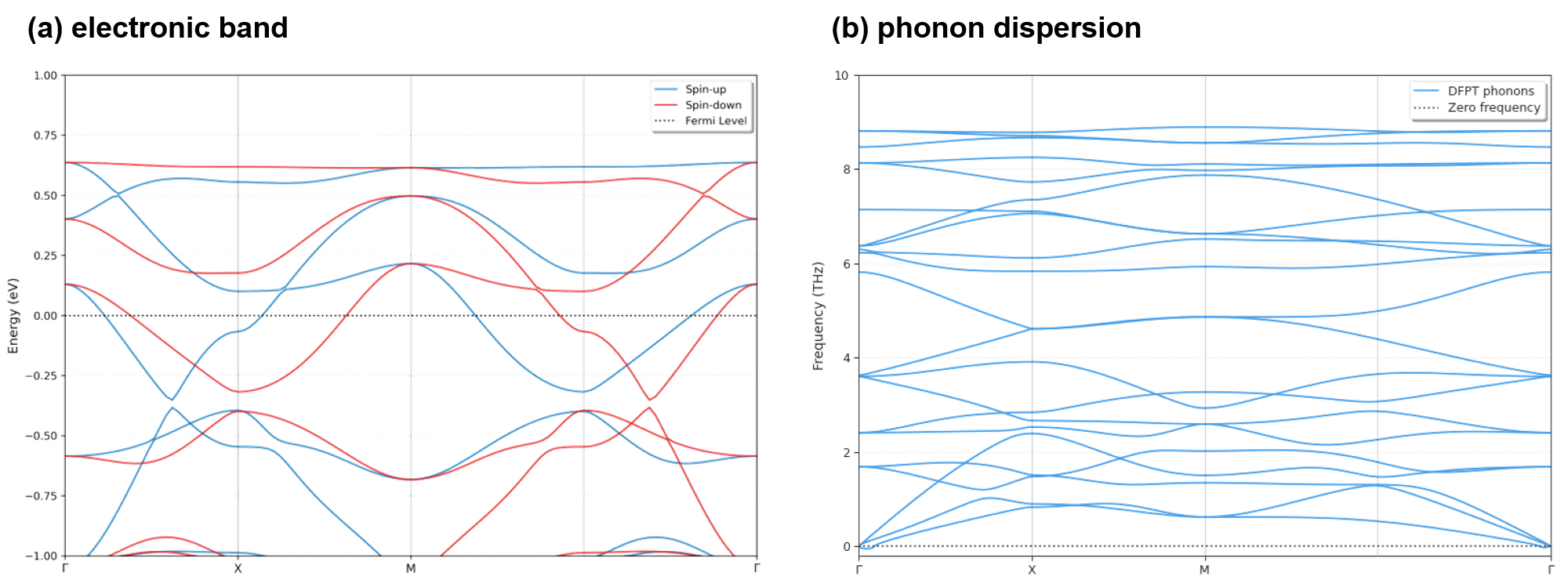}
  \caption{\textbf{Electronic band structure and phonon dispersion of the representative candidate TaCo$_2$Se$_4$.} (a) Spin-resolved DFT bands; different colors denote the two spin channels, and the dashed line indicates the Fermi level. (b) VASP-DFPT phonon dispersion of the reference structure. Force constants are obtained using a $2\times2\times1$ supercell and post-processed with phonopy along $\Gamma$-X-M-$\Gamma$}
  \label{fig:lieb_results}
\end{figure}

TaCo$_2$Se$_4$ illustrates this analysis. In the upper-right schematic of Figure~\ref{fig:lieb_workflow}, Ta occupies the cell corner, Co occupies two edge-center sites, and Se occupies four ligand sites, forming a Lieb-lattice-derived metal framework and coordination environment. On the reference relaxed geometry, the tested AFM configuration has an energy approximately $14.86\ \mathrm{meV/atom}$ below the FM-like reference, with self-consistent local moments of approximately $-1.237$ and $+1.237\ \mu_{\mathrm{B}}$ on the two Co sites. Together with the spin-resolved bands in Figure~\ref{fig:lieb_results}(a), these results support its selection as an altermagnetic candidate.

We also perform a VASP-DFPT phonon check for a reference structure of TaCo$_2$Se$_4$. The phonon spectrum in Figure~\ref{fig:lieb_results}(b) calculates along $\Gamma$-X-M-$\Gamma$.  Calculation settings are provided in Methods.

\FloatBarrier

\section{Discussion}

UniMatSim serves as a lightweight execution layer for UMLIP-assisted materials screening. ASE and pymatgen provide foundational capabilities for structures, calculators, and materials analysis, while general workflow infrastructures such as AiiDA and atomate2 support broader job management, database provenance, and DFT and machine learning force-field workflows\cite{huber2022automated,rosen2024atomate2}. UniMatSim complements these tools by focusing on the repeated execution of selected screening protocols after model selection. Users provide a workflow configuration, material dataset, and potential model through the CLI. The framework manages model initialization, task dependencies, propagation of updated structures, batch materials processing, runtime state, and candidate aggregation. Figure~\ref{fig:architecture} illustrates this configuration-driven execution path.

The unified potential model interface enables the same task graph to be reused across supported backends without model-specific initialization code in individual research scripts. For models deployed in separate software environments, the REST API and \lstinline{RemoteCalculator} provide remote access to energy, force, and stress evaluations through the ASE calculator interface.

The current implementation records runtime state through the file system, material-level outputs, and checkpoint/manifest files, keeping calculation records directly inspectable across repeated screening runs. Integration with database-backed workflow infrastructure can extend provenance tracking, multiuser concurrency, and cluster-level resource orchestration. For two-dimensional materials, UniMatSim incorporates in-plane cell constraints, $(n,n,1)$ supercells, and two-dimensional Brillouin-zone path handling into the relevant tasks, reducing repeated implementation of low-dimensional calculation rules across projects.

Future development of UniMatSim will focus on: (1) extending support to emerging UMLIPs; (2) constructing active learning workflows that close the loop between machine learning potential evaluation, DFT validation, and model optimization; (3) developing a terminal user interface (TUI) based on Textual to improve usability for users without programming experience; and (4) integrating database backends, such as MongoDB, to improve management of large-scale simulation data.

We present UniMatSim, a configuration-driven workflow framework for UMLIP-assisted high-throughput materials screening. Its core functionality binds workflow configurations, material data, and potential models or model checkpoints into repeatable multistage calculations and manages task dependencies, computational backends, propagation of updated structures, batch materials processing, checkpointing and resumption, and candidate aggregation. The unified potential model interface enables registered tasks to be reused across supported backends. The REST API provides an independent path for remote potential model invocation, while lifecycle management of complete workflows remains the responsibility of the workflow layer.

The two-dimensional Lieb-lattice application illustrates this execution model at screening scale: the same optimization--elastic--phonon protocol is executed across multiple model instances. The resulting candidate sets are compared and integrated using consistent material identifiers, yielding 393 consensus candidates for subsequent first-principles analysis. Future work will extend potential model backends, data management, and monitoring of long-running tasks, and explore configurable iterative workflows combining DFT validation, training-data preparation, and model updates.

\section{Methods}

\subsection{Lieb-lattice candidate generation}

The initial library was constructed by site-specific elemental substitution of a WFe$_2$Te$_4$ Lieb-lattice template. The W site defines $M$, the two Fe sites define $M'$, and the four Te sites define $X$ in $MM'_2X_4$. In the ideal square template, $M$ occupies the cell corner at fractional coordinates $(0,0,0.5)$, while the two symmetry-related $M'$ sites occupy edge centers at $(0,0.5,0.5)$ and $(0.5,0,0.5)$. The $M$ pool contains 29 elements: Ag, Au, Cd, Co, Cr, Cu, Fe, Hf, Hg, Ir, Mn, Mo, Nb, Ni, Os, Pd, Pt, Re, Rh, Ru, Sc, Ta, Tc, Ti, V, W, Y, Zn, and Zr. The $M'$ pool contains Fe, Co, Ni, Cd, Mn, and V, and the $X$ pool contains O, S, Se, Te, N, P, and As. For each candidate, both Fe-derived sites were replaced by the same $M'$ element, and all four Te-derived sites were replaced by the same $X$ element. Combinations with $M=M'$ were excluded. Because all six $M'$ elements belong to the 29-element $M$ pool, this procedure yields $6\times28\times7=1{,}176$ distinct compositions.

\subsection{UMLIP-based structural, elastic, and phonon screening}

The four Uni2D checkpoints were loaded through the MatterSim backend\cite{2dmodel,yang2024mattersim}. Structures were relaxed with QuasiNewton  using a maximum-force threshold of $f_{\mathrm{max}}=0.01\ \mathrm{eV}/\text{\AA}$ and a maximum of 500 optimization steps. Cell relaxation used the \texttt{[1,1,0,0,0,1]} mask, allowing in-plane cell degrees of freedom while keeping the vacuum direction fixed. Axial strains $[-0.01,-0.005,0.005,0.01]$ and shear strains $[-0.06,-0.03,0.03,0.06]$ were applied to calculate the elastic stiffness tensor; Elastic constants were evaluated using the optimized equilibrium structures. Elastic stability was assessed using Born--Huang-type conditions for the in-plane stiffness tensor\cite{born1954dynamical,mouhat2014necessary,mech2d}: square lattices were required to satisfy $C_{11}+C_{12}>0$, $C_{11}-C_{12}>0$, and $C_{66}>0$; rectangular lattices were required to satisfy $C_{11}>0$, $C_{22}>0$, $C_{66}>0$, and $C_{11}C_{22}-C_{12}^2>0$; for oblique lattices, all eigenvalues of the in-plane stiffness submatrix were required to be positive. Phonon calculations used the finite-displacement method\cite{togo2015first}, a $5\times5\times1$ supercell, an atomic displacement amplitude of 0.05~\AA, and 100 sampling points along the high-symmetry paths of the two-dimensional Brillouin zone. Screening applied energy thresholds of $-0.005$ eV at high-symmetry points and $-0.005$ eV along connecting path segments. The same optimization--elastic--phonon protocol was run for Uni2D instances trained with random seeds 42, 123, 2025, and 8888.

\subsection{DFT calculations}
\label{sec:dft-calculations}

Density functional theory (DFT) calculations were performed using the Vienna Ab initio Simulation Package (VASP)\cite{vasp,vasp2} and the projector augmented-wave (PAW) method\cite{kresse1999paw}. Structural relaxations and static self-consistent-field calculations used a plane-wave cutoff of 520 eV, \texttt{PREC} = \texttt{Accurate}, and spin polarization with \texttt{ISPIN} = 2. Relaxations used \texttt{EDIFF} = $1\times10^{-5}$ eV, \texttt{EDIFFG} = $-0.01$ eV/\AA, \texttt{IBRION} = 2. Relaxation and static calculations used a $\Gamma$-centered $13\times13\times1$ $k$-point mesh, with \texttt{ISMEAR} = $-5$ and \texttt{SIGMA} = 0.05 eV. Static calculations used \texttt{GGA} = \texttt{Bo} and \texttt{LUSE\_VDW} = \texttt{True} to obtain energies and charge densities.

Reference geometries were obtained from spin-polarized relaxations initialized with FM-like magnetic configurations. AFM configurations were generated by reversing selected initial atomic magnetic moments and evaluated through spin-polarized single-point calculations on the reference geometries. The energy difference was defined as $\Delta E=(E_{\mathrm{AFM}}-E_{\mathrm{FM\text{-}like}})/N$, where $N$ is the number of atoms. Configurations satisfying $\Delta E\leq0.001\ \mathrm{eV/atom}$ were examined using the self-consistent local magnetic moments. Initial magnetic assignments and converged local moments were treated separately in the analysis.

\subsection{\texorpdfstring{DFT phonon calculation for TaCo$_2$Se$_4$}{DFT phonon calculation for TaCo2Se4}}

Density functional perturbation theory calculations for TaCo$_2$Se$_4$ were performed in VASP. Relaxation of the reference structure used a $4\times4\times1$ Monkhorst--Pack $k$-point mesh. Force constants were calculated using a $2\times2\times1$ supercell containing 28 atoms, a $2\times2\times1$ Monkhorst--Pack $k$-point mesh. Force constants were subsequently post-processed using phonopy 2.47.1\cite{togo2015first,phono3py} to calculate the phonon dispersion along $\Gamma$--X--M--$\Gamma$.

\subsection{Computational environment}

The high-throughput screening runs were performed on an HPC cluster managed by SLURM. Compute nodes were equipped with two AMD EPYC 9A14 processors, each with 96 physical cores. The Uni2D checkpoints were loaded through the MatterSim backend.

\subsection{Use of generative AI tools}

During preparation of this manuscript, the authors used GPT-5 to improve readability and coherence and Claude 4.5 to assist with code optimization.

\section*{Data availability}
The dataset supporting the findings of this study is publicly available in Zenodo at \url{https://doi.org/10.5281/zenodo.23854427}.

\section*{Code availability}
The UniMatSim source code is available at \url{https://gitee.com/haidi-hfut/unimatsim} under the MIT License.

\section*{Acknowledgments}
UniMatSim builds on several major open-source projects, including pymatgen for materials analysis, ASE for atomistic simulations, phono3py for phonon calculations, and the original implementations of machine learning interatomic potentials such as CHGNet, M3GNet, and MACE. We acknowledge the early work of the matcalc package, which provided valuable inspiration for the methodological objectives of this research. We also thank Dr. Jielan Li at Microsoft for valuable contributions to the Boltzmann transport equation (BTE) code.
This work was supported by the robotic AI-Scientist platform of the Chinese Academy of Sciences, the Open Research Fund of the State Key Laboratory of Precision and Intelligent Chemistry, and the National Natural Science Foundation of China (22203026), the National Natural Science Foundation of China (22503093, 42550106, 22288201, 22173093), the Laoshan Laboratory (LSKJ202300305), the Innovation Program for Quantum Science and Technology (2021ZD0303306), the Strategic Priority Research Program of the Chinese Academy of Sciences (XDB0450101), by the Anhui Provincial Key Research and Development Program (2022a05020052), the National Key Research and Development Program of China (2021YFB0300600), the Anhui Province Science and Technology Innovation Project (202423k09020010), and the CAS Project for Young Scientists in Basic Research (YSBR-005). The numerical calculations in this paper have been done on the supercomputing system in the Supercomputing Center of University of Science and Technology of China. 

\section*{Author contributions}
Y.X. developed and implemented the UniMatSim software and contributed to writing the manuscript. Y.N. , W.W and Y.G. contributed to software testing. H.W. and W.H. supervised the research and contributed to the review and revision of the manuscript. All authors read and approved the final version of the manuscript.

\section*{Corresponding authors}
Correspondence to Haidi Wang and Wei Hu.

\section*{Competing interests}
The authors declare no competing interests.

\bibliographystyle{unsrtnat}
\bibliography{references}


\clearpage

\setcounter{page}{1}
\renewcommand{\thepage}{S\arabic{page}}
\setcounter{figure}{0}
\renewcommand{\thefigure}{S\arabic{figure}}
\setcounter{table}{0}
\renewcommand{\thetable}{S\arabic{table}}
\setcounter{equation}{0}
\renewcommand{\theequation}{S\arabic{equation}}
\setcounter{section}{0}
\renewcommand{\thesection}{S\arabic{section}}
\setcounter{subsection}{0}

\renewcommand{\theHsection}{SI.\arabic{section}}
\renewcommand{\theHfigure}{SI.\arabic{figure}}
\renewcommand{\theHtable}{SI.\arabic{table}}
\renewcommand{\theHequation}{SI.\arabic{equation}}

\pdfbookmark[0]{Supplementary Information}{supplementary-information}
\begin{center}
{\Large\bfseries Supplementary Information}\\[0.5em]
{\large UniMatSim: An automated framework for high-throughput materials simulations using universal machine learning interatomic potentials}\\[1em]
{Yanjin Xiang$^{1}$, Yihan Nie$^{1}$, Weiyi Wang$^{3}$ and Yunzhi Gao$^{1}$\\[0.25em]
Haidi Wang$^{2,*}$ and Wei Hu$^{3,*}$}\\[0.65em]
{\small $^*$Corresponding authors: haidi@hfut.edu.cn (Haidi Wang), whuustc@ustc.edu.cn (Wei Hu)}
\end{center}

\clearpage

\section{Consistency of candidate sets across models trained with different random seeds}

This study uses four Uni2D model instances trained with different random seeds (42, 123, 2025, and 8888), all loaded through the MatterSim backend. Each instance independently performs structural optimization, elastic stability screening, and phonon screening on the same initial Lieb-lattice candidate set, yielding 438, 438, 428, and 449 candidates, respectively. The four instances share the same Uni2D architecture and differ in their training random seeds.

\begin{figure}[htbp]
\centering
\includegraphics[width=0.9\textwidth]{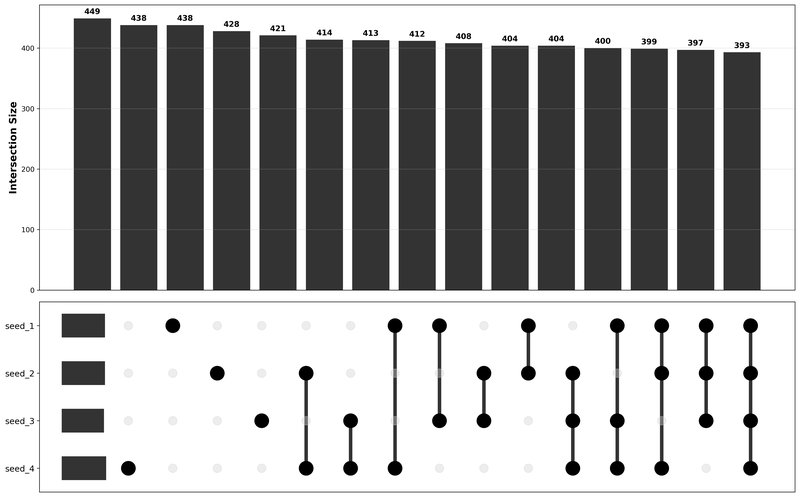}
\caption{\textbf{UpSet plot of four model candidate sets generated by UniMatSim.} UniMatSim's analysis module calculates intersections among the phonon-screened candidate sets from four Uni2D models and generates this plot. All model checkpoints are loaded through the MatterSim backend. Labels \texttt{seed\_1}--\texttt{seed\_4} correspond to training random seeds 42, 123, 2025, and 8888, respectively. The upper bars indicate candidate counts in sets or intersections; black dots and connecting lines below identify the model combinations involved, and horizontal bars on the left indicate individual set sizes. Intersection counts include candidates also retained by other models, so different bars may overlap in membership and cannot be summed directly. The strict four-model intersection contains 393 candidates used for subsequent DFT screening.}
\label{fig:upset_si}
\end{figure}

Consistency among the four candidate sets passing phonon screening is analyzed using UniMatSim's analysis module, which generates the UpSet plot in Figure~\ref{fig:upset_si}. The strict four-model intersection contains 393 candidate structures used as inputs for subsequent first-principles calculations.

\FloatBarrier

\section{UniMatSim command-line and workflow configuration examples}
\label{sec:si-cli}

This section provides minimal runnable examples of UniMatSim's configuration-driven usage. For multistage UMLIP prescreening, the core user inputs are a workflow configuration, material data, and a potential model; the CLI binds them into a resumable run. The examples show how to generate material JSON from raw structures, record the screening protocol in YAML, and launch the complete multistage UMLIP prescreening workflow with a single \texttt{workflow run} command.

\subsection{Material data import}

When raw structures are distributed across directories as POSCAR/CONTCAR or \texttt{.vasp} files, data import commands can first generate JSON inputs readable by \texttt{calc} or \texttt{workflow}. For structure libraries with nested directories, \texttt{--merge} generates a merged structure file. The structure metadata fields \texttt{material\_name}, \texttt{stacking\_mode}, and \texttt{order} are used for subsequent output naming and provenance.

\begin{lstlisting}
unimatsim data import poscar -i after_relax -O imported_structures --pattern "**/POSCAR" --merge
\end{lstlisting}

\subsection{Single-task CLI examples}

Individual calculation tasks can be executed directly through \texttt{calc}. In \texttt{calc}, \texttt{-o}/\texttt{--output} specifies the output directory. Without explicit format flags, the command writes all formats supported by the task; when format flags are supplied, only the requested outputs are retained. For example, the second command below saves only JSON and figure outputs.

\begin{lstlisting}
unimatsim calc optimize -i Si.cif -m chgnet -o optimize_results
unimatsim calc optimize -i Si.cif -m chgnet --json --figure -o optimize_results
unimatsim calc phonon -i monolayer.vasp -m mattersim --is-2d y --supercell 4,4,1 -o phonon_2d
\end{lstlisting}

\subsection{Minimal material JSON input}

Batch inputs generated by \texttt{data import poscar --merge} store pymatgen-compatible Structure dictionaries in a top-level \texttt{structures} list. The workflow expands each structure in this list into a separate material run. The following example illustrates the input structure using a single material.

\begin{lstlisting}[language=json,caption={Minimal materials JSON example that can be used directly as a workflow batch structure input.},label={lst:materials-json},captionpos=t]
{
  "@module": "pymatgen.core.structure",
  "@class": "StructureList",
  "structures": [
    {
      "@module": "pymatgen.core.structure",
      "@class": "Structure",
      "charge": 0,
      "lattice": {
        "matrix": [
          [5.43, 0.0, 0.0],
          [0.0, 5.43, 0.0],
          [0.0, 0.0, 20.0]
        ],
        "pbc": [true, true, true]
      },
      "properties": {
        "material_name": "Si",
        "stacking_mode": "unknown",
        "order": "unknown",
        "material_label": "Si"
      },
      "sites": [
        {
          "species": [{"element": "Si", "occu": 1}],
          "abc": [0.0, 0.0, 0.5],
          "properties": {},
          "label": "Si"
        }
      ]
    }
  ],
  "total_structures": 1
}
\end{lstlisting}

\subsection{Complete Lieb-lattice workflow YAML}

For multistage screening, YAML or JSON files serve both as records of the scientific protocol and as execution configurations. The YAML example below illustrates a reusable configuration for the Lieb-lattice screening workflow, incorporating the computational tasks and parameters documented for this study.: structures undergo optimization with two-dimensional handling, elastic screening, and finally a finite-displacement phonon check. The model type and checkpoint are also specified in metadata. Task order, physical parameters, screening thresholds, and model inputs are thus stored in a single reusable configuration without a separate project-specific driver script.

\begin{lstlisting}[language=yaml]
metadata:
  description: Lieb screening with explicit parameters from the historical scripts
  ml_potential: mattersim
  ml_params:
    model_path: ./models/best_model.pth
    device: cpu
  potential_type: ml
  privilege: material
  stable_material: true
  summary_mode: aggregate
  individual_task_summary: true

tasks:
  - name: opt
    type: optimization
    config:
      optimizer: QuasiNewton
      fmax: 0.01
      steps: 500
      is_2d: true
      relax_cell: true
      cell_filter_mask: [1, 1, 0, 0, 0, 1]
      save_structure: true

  - name: symmetry
    type: symmetry
    config:
      target_spacegroups: [111]

  - name: elastic
    type: elastic
    config:
      fmax: 0.1
      use_equilibrium: true
      check_stability: true
      cell_filter_mask: [1, 1, 0, 0, 0, 1]

  - name: phonon
    type: phonon
    config:
      supercell: [5, 5, 1]
      delta: 0.05
      is_2d: true
      check_stability: true
      spoints_threshold: -0.005
      gpoints_threshold: -0.005
      path_generation_mode: manual
      custom_bz_path: GXMG
      filter_2d_bz_path: false
      band_npoints: 100
      dos_kpts: [20, 20, 1]
      dos_npts: 100
      dos_width: 0.001
      emax: 0.035
      save_bandstructure: true
      save_dos: true
      save_phonon_plot: true

dependencies:
  opt: symmetry
  symmetry: elastic
  elastic: phonon
  phonon: null

\end{lstlisting}

The YAML listing provides a reusable configuration for the Lieb-lattice screening workflow, including the material input, potential backend, task parameters, screening thresholds, and task dependencies. 

\subsection{Workflow launch, resumption, and lifecycle}

Once the workflow file and material JSON are prepared, the complete multistage UMLIP prescreening can be launched directly through \texttt{workflow}. For \texttt{workflow}, \texttt{-O}/\texttt{--output-dir} overrides the output directory, while checkpointing is controlled by the runtime options \texttt{--id} and \texttt{--no-checkpoint}. Internally, \texttt{workflow run} handles material iteration, execution of task dependencies, propagation of updated structures, and candidate aggregation according to the configuration. Upstream task results and execution states are retained in the workflow state, making the commands below the user-facing entry point for the configuration-driven prescreening presented here. When a consistent \texttt{--id} is supplied, the parent checkpoint/manifest records the original configuration, inputs, output directory, model, and child workflow identifiers. After an interruption, unfinished structures can be resumed using only \texttt{resume --id}.

\begin{lstlisting}
unimatsim workflow run -C workflow_2d.yaml -i material.json --id my_workflow
unimatsim workflow run -C workflow_2d.yaml -i material.json -O workflow_results --id my_workflow
unimatsim workflow resume --id my_workflow
\end{lstlisting}

\section{Crystallographic parameters of selected Lieb-lattice candidates}

This section reports crystallographic parameters for the 59 selected Lieb-lattice candidates, including optimized lattice constants, unit-cell volumes, and crystallographic space groups. All data were obtained from the final structures relaxed using VASP. These materials form two-dimensional layered Lieb-lattice structures with vacuum along the $c$ axis. Variations in lattice parameters reflect differences in chemical composition. Space-group analysis indicates highly consistent symmetry: 58 of the 59 compounds belong to space group P-42m, with Fe$_2$Te$_4$Ru as the sole exception, belonging to Cmm2.

\begin{longtable}{c c c c c c}
\caption{Basic crystallographic parameters of 59 selected Lieb-lattice candidates}\\
\toprule
No. & Formula & Lattice parameters & Volume & Atoms & Space group \\
\midrule
\endfirsthead

\caption[]{Basic crystallographic parameters of 59 selected Lieb-lattice candidates (continued)}\\
\toprule
No. & Formula & Lattice parameters & Volume & Atoms & Space group \\
\midrule
\endhead

\bottomrule
\endfoot

1 & \ce{Co2CuSe4} & $a=5.402$, $b=5.402$, $c=19.985$ & 583.16 & 7 & P-42m \\
2 & \ce{Co2WSe4} & $a=5.487$, $b=5.487$, $c=19.985$ & 601.69 & 7 & P-42m \\
3 & \ce{CrFe2S4} & $a=5.257$, $b=5.257$, $c=19.985$ & 552.23 & 7 & P-42m \\
4 & \ce{CrFe2Se4} & $a=5.416$, $b=5.416$, $c=19.985$ & 586.31 & 7 & P-42m \\
5 & \ce{Fe2CuTe4} & $a=5.246$, $b=5.246$, $c=19.985$ & 550.02 & 7 & P-42m \\
6 & \ce{Fe2MoSe4} & $a=5.417$, $b=5.417$, $c=19.985$ & 586.45 & 7 & P-42m \\
7 & \ce{Fe2NiS4} & $a=5.039$, $b=5.039$, $c=19.985$ & 507.42 & 7 & P-42m \\
8 & \ce{Fe2NiSe4} & $a=5.278$, $b=5.278$, $c=19.985$ & 556.68 & 7 & P-42m \\
9 & \ce{Fe2OsSe4} & $a=5.325$, $b=5.325$, $c=19.985$ & 566.71 & 7 & P-42m \\
10 & \ce{Fe2PtSe4} & $a=5.359$, $b=5.359$, $c=19.985$ & 574.01 & 7 & P-42m \\
11 & \ce{Fe2ReSe4} & $a=5.412$, $b=5.412$, $c=19.985$ & 585.40 & 7 & P-42m \\
12 & \ce{Fe2ReTe4} & $a=5.542$, $b=5.542$, $c=19.985$ & 613.85 & 7 & P-42m \\
13 & \ce{Fe2TcP4} & $a=5.446$, $b=5.446$, $c=19.985$ & 592.80 & 7 & P-42m \\
14 & \ce{Fe2TcS4} & $a=5.289$, $b=5.288$, $c=19.985$ & 558.94 & 7 & P-42m \\
15 & \ce{Fe2TcSe4} & $a=5.419$, $b=5.419$, $c=19.985$ & 586.90 & 7 & P-42m \\
16 & \ce{Fe2TcTe4} & $a=5.595$, $b=5.595$, $c=19.985$ & 625.54 & 7 & P-42m \\
17 & \ce{Fe2Te4Pt} & $a=5.309$, $b=5.309$, $c=19.985$ & 563.38 & 7 & P-42m \\
18 & \ce{Fe2Te4Rh} & $a=5.574$, $b=5.574$, $c=19.985$ & 620.91 & 7 & P-42m \\
19 & \ce{Fe2Te4Ru} & $a=5.551$, $b=5.551$, $c=19.985$ & 615.46 & 7 & Cmm2 \\
20 & \ce{Fe2WSe4} & $a=5.556$, $b=5.556$, $c=19.985$ & 616.90 & 7 & P-42m \\
21 & \ce{HfMn2S4} & $a=6.123$, $b=6.123$, $c=19.985$ & 749.16 & 7 & P-42m \\
22 & \ce{HfMn2Se4} & $a=6.272$, $b=6.272$, $c=19.985$ & 786.15 & 7 & P-42m \\
23 & \ce{HfMn2Te4} & $a=6.503$, $b=6.503$, $c=19.985$ & 845.06 & 7 & P-42m \\
24 & \ce{Mn2CrSe4} & $a=5.852$, $b=5.852$, $c=19.985$ & 684.39 & 7 & P-42m \\
25 & \ce{Mn2CrTe4} & $a=6.082$, $b=6.082$, $c=19.985$ & 739.32 & 7 & P-42m \\
26 & \ce{Mn2CuS4} & $a=5.477$, $b=5.477$, $c=19.985$ & 599.54 & 7 & P-42m \\
27 & \ce{Mn2MoS4} & $a=5.654$, $b=5.654$, $c=19.985$ & 638.94 & 7 & P-42m \\
28 & \ce{Mn2MoSe4} & $a=5.778$, $b=5.778$, $c=19.985$ & 667.16 & 7 & P-42m \\
29 & \ce{Mn2NbS4} & $a=5.828$, $b=5.828$, $c=19.985$ & 678.84 & 7 & P-42m \\
30 & \ce{Mn2NbSe4} & $a=5.982$, $b=5.982$, $c=19.985$ & 715.13 & 7 & P-42m \\
31 & \ce{Mn2NbTe4} & $a=6.212$, $b=6.212$, $c=19.985$ & 771.17 & 7 & P-42m \\
32 & \ce{Mn2NiTe4} & $a=5.412$, $b=5.412$, $c=19.985$ & 585.36 & 7 & P-42m \\
33 & \ce{Mn2OsS4} & $a=5.280$, $b=5.280$, $c=19.985$ & 557.21 & 7 & P-42m \\
34 & \ce{Mn2OsSe4} & $a=5.377$, $b=5.377$, $c=19.985$ & 577.80 & 7 & P-42m \\
35 & \ce{Mn2ReSe4} & $a=5.571$, $b=5.572$, $c=19.985$ & 620.36 & 7 & P-42m \\
36 & \ce{Mn2RuS4} & $a=5.290$, $b=5.286$, $c=19.985$ & 558.83 & 7 & P-42m \\
37 & \ce{Mn2RuSe4} & $a=5.430$, $b=5.430$, $c=19.985$ & 589.31 & 7 & P-42m \\
38 & \ce{Mn2TcSe4} & $a=5.580$, $b=5.580$, $c=19.985$ & 622.37 & 7 & P-42m \\
39 & \ce{Mn2TcTe4} & $a=5.790$, $b=5.790$, $c=19.985$ & 670.06 & 7 & P-42m \\
40 & \ce{Mn2Te4Mo} & $a=5.970$, $b=5.970$, $c=19.985$ & 712.29 & 7 & P-42m \\
41 & \ce{Mn2Te4Pd} & $a=5.482$, $b=5.482$, $c=19.985$ & 600.50 & 7 & P-42m \\
42 & \ce{Mn2VS4} & $a=5.670$, $b=5.670$, $c=19.985$ & 642.53 & 7 & P-42m \\
43 & \ce{Mn2VSe4} & $a=5.854$, $b=5.854$, $c=19.985$ & 684.94 & 7 & P-42m \\
44 & \ce{Mn2VTe4} & $a=6.099$, $b=6.099$, $c=19.985$ & 743.35 & 7 & P-42m \\
45 & \ce{Mn2WS4} & $a=5.656$, $b=5.656$, $c=19.985$ & 639.33 & 7 & P-42m \\
46 & \ce{TaCo2Se4} & $a=5.653$, $b=5.653$, $c=19.985$ & 638.69 & 7 & P-42m \\
47 & \ce{TaCo2Te4} & $a=5.873$, $b=5.873$, $c=19.985$ & 689.40 & 7 & P-42m \\
48 & \ce{TaMn2S4} & $a=5.837$, $b=5.836$, $c=19.985$ & 680.82 & 7 & P-42m \\
49 & \ce{TaMn2Se4} & $a=5.981$, $b=5.981$, $c=19.985$ & 714.89 & 7 & P-42m \\
50 & \ce{TiCo2S4} & $a=5.445$, $b=5.445$, $c=19.985$ & 592.53 & 7 & P-42m \\
51 & \ce{TiCo2Se4} & $a=5.602$, $b=5.600$, $c=19.985$ & 626.92 & 7 & P-42m \\
52 & \ce{TiFe2S4} & $a=5.614$, $b=5.614$, $c=19.985$ & 629.87 & 7 & P-42m \\
53 & \ce{TiMn2S4} & $a=5.930$, $b=5.930$, $c=19.985$ & 702.87 & 7 & P-42m \\
54 & \ce{TiMn2Te4} & $a=6.343$, $b=6.343$, $c=19.985$ & 804.11 & 7 & P-42m \\
55 & \ce{VCo2S4} & $a=5.310$, $b=5.310$, $c=19.985$ & 563.40 & 7 & P-42m \\
56 & \ce{VFe2S4} & $a=5.243$, $b=5.242$, $c=19.985$ & 549.31 & 7 & P-42m \\
57 & \ce{ZnFe2S4} & $a=5.353$, $b=5.353$, $c=19.985$ & 572.62 & 7 & P-42m \\
58 & \ce{ZnFe2Se4} & $a=5.451$, $b=5.451$, $c=19.985$ & 593.78 & 7 & P-42m \\
59 & \ce{ZrMn2Te4} & $a=6.519$, $b=6.519$, $c=19.985$ & 849.34 & 7 & P-42m \\
\addlinespace[0.75ex]
\multicolumn{6}{@{}p{0.94\textwidth}@{}}{\footnotesize\textit{Note:} Fe$_2$RuTe$_4$ (listed as Fe$_2$Te$_4$Ru above) adopts a Cmm2 reference structure and is retained for its spin-resolved band features. Its magnetic-symmetry assignment remains open.} \\

\end{longtable}

\section{Spin-resolved band structures of selected Lieb-lattice candidates}
\label{sec:si-band-structures}

This section presents spin-resolved DFT band structures for the 59 selected Lieb-lattice candidates from the magnetic and electronic screening of 393 UMLIP consensus candidates. The consensus set is the strict intersection of screening results from four Uni2D models trained with different random seeds (42, 123, 2025, and 8888), all loaded through the MatterSim backend. The band structures were calculated using spin-polarized DFT to examine spin-dependent band features. They are presented as candidate-level electronic-structure results rather than definitive magnetic-symmetry classifications. The materials are grouped by chemical composition for comparison.

\begin{figure}[H]
\centering
\begin{minipage}{0.48\textwidth}
    \centering
    \includegraphics[width=\textwidth]{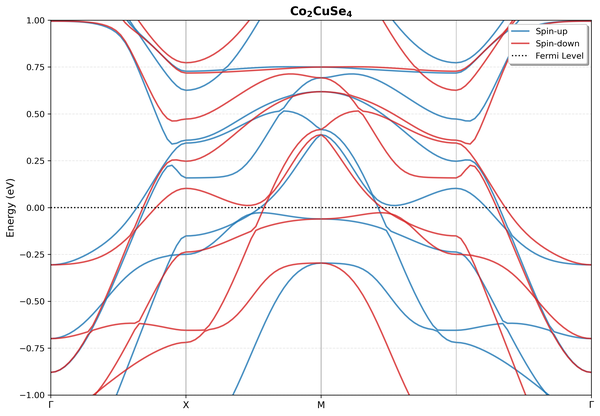}
    \caption*{\small Co$_2$CuSe$_4$ }
\end{minipage}
\hfill
\begin{minipage}{0.48\textwidth}
    \centering
    \includegraphics[width=\textwidth]{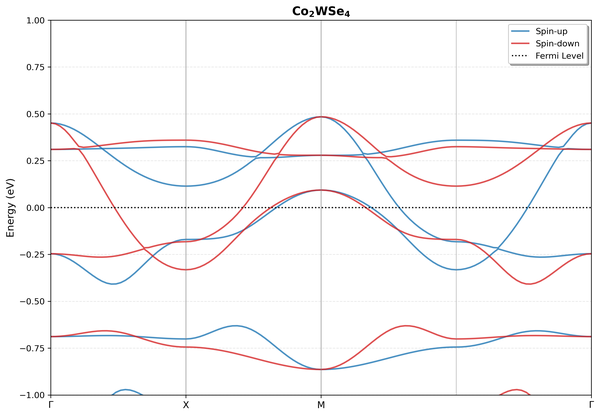}
    \caption*{\small Co$_2$WSe$_4$ }
\end{minipage}
\\[1em]
\begin{minipage}{0.48\textwidth}
    \centering
    \includegraphics[width=\textwidth]{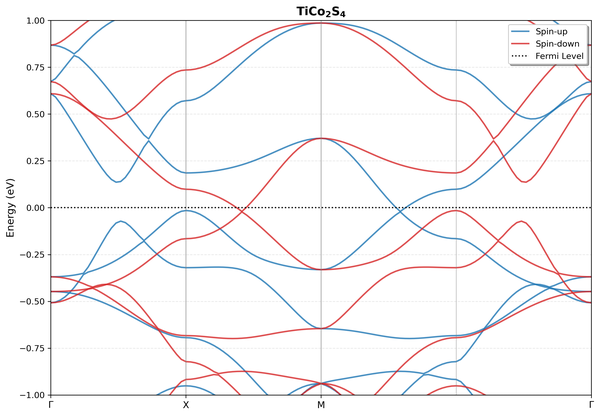}
    \caption*{\small TiCo$_2$S$_4$ }
\end{minipage}
\hfill
\begin{minipage}{0.48\textwidth}
    \centering
    \includegraphics[width=\textwidth]{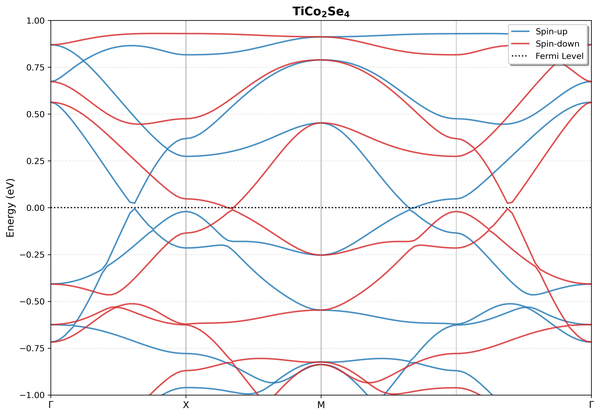}
    \caption*{\small TiCo$_2$Se$_4$ }
\end{minipage}
\\[1em]
\begin{minipage}{0.48\textwidth}
    \centering
    \includegraphics[width=\textwidth]{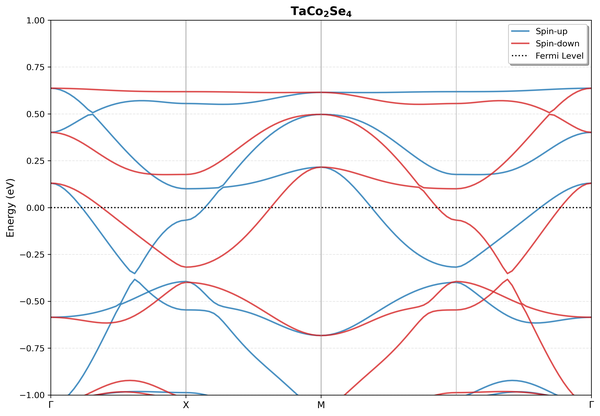}
    \caption*{\small TaCo$_2$Se$_4$ }
\end{minipage}
\hfill
\begin{minipage}{0.48\textwidth}
    \centering
    \includegraphics[width=\textwidth]{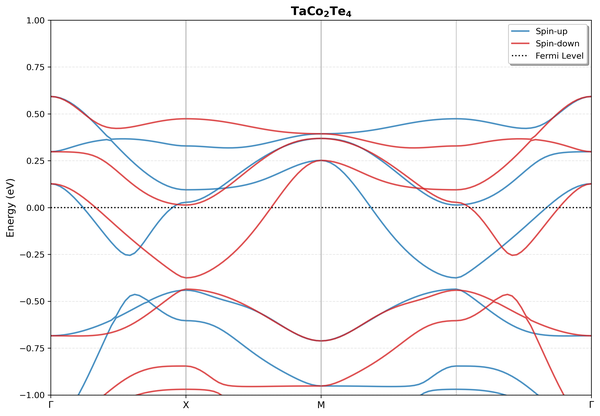}
    \caption*{\small TaCo$_2$Te$_4$ }
\end{minipage}
\end{figure}

\begin{figure}[H]
\centering
\begin{minipage}{0.48\textwidth}
    \centering
    \includegraphics[width=\textwidth]{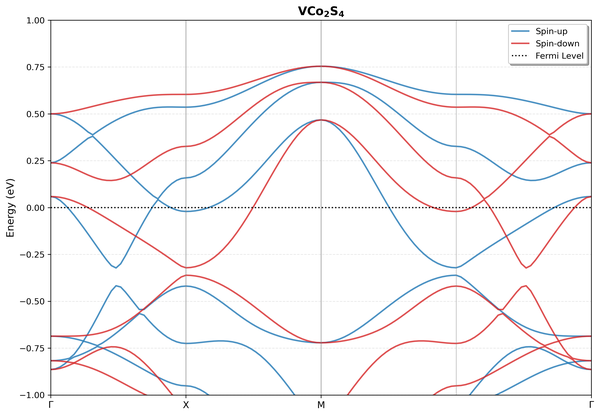}
    \caption*{\small VCo$_2$S$_4$ }
\end{minipage}
\end{figure}

\clearpage


\begin{figure}[H]
\centering
\begin{minipage}{0.48\textwidth}
    \centering
    \includegraphics[width=\textwidth]{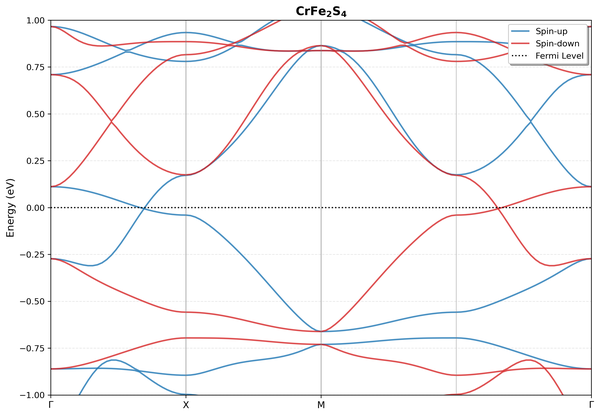}
    \caption*{\small CrFe$_2$S$_4$ }
\end{minipage}
\hfill
\begin{minipage}{0.48\textwidth}
    \centering
    \includegraphics[width=\textwidth]{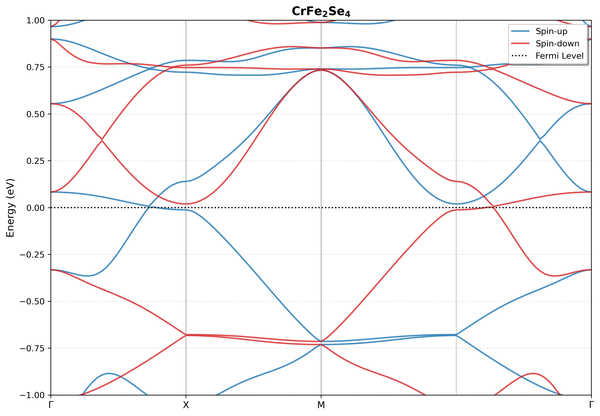}
    \caption*{\small CrFe$_2$Se$_4$ }
\end{minipage}
\end{figure}

\clearpage


\begin{figure}[H]
\centering
\begin{minipage}{0.48\textwidth}
    \centering
    \includegraphics[width=\textwidth]{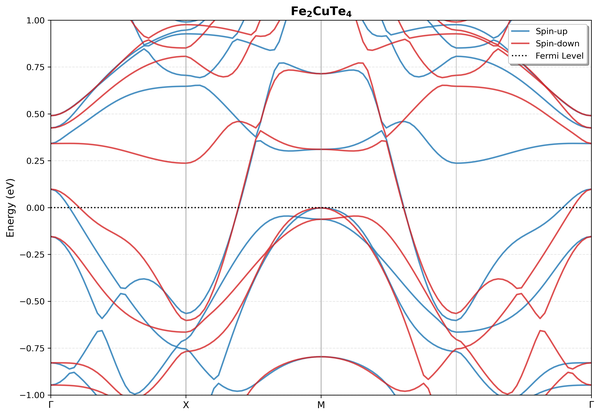}
    \caption*{\small Fe$_2$CuTe$_4$ }
\end{minipage}
\hfill
\begin{minipage}{0.48\textwidth}
    \centering
    \includegraphics[width=\textwidth]{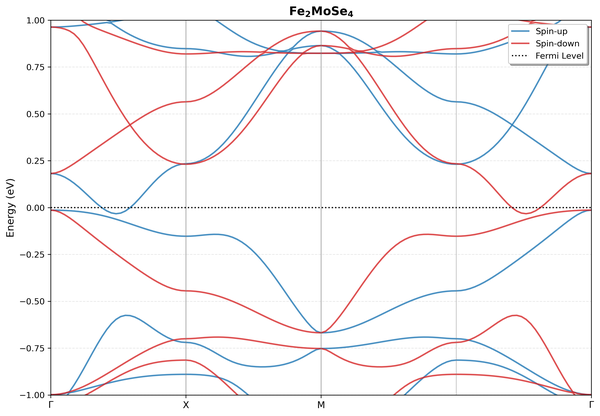}
    \caption*{\small Fe$_2$MoSe$_4$ }
\end{minipage}
\\[1em]
\begin{minipage}{0.48\textwidth}
    \centering
    \includegraphics[width=\textwidth]{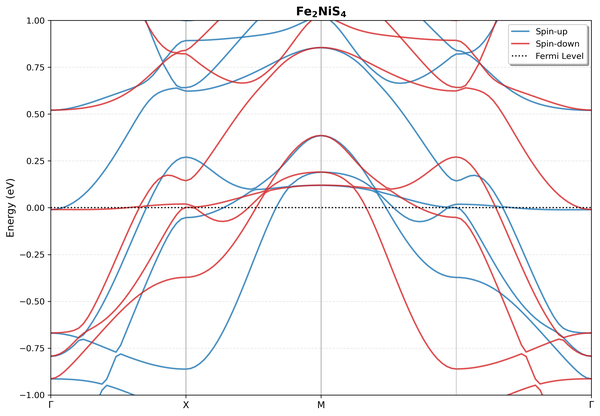}
    \caption*{\small Fe$_2$NiS$_4$ }
\end{minipage}
\hfill
\begin{minipage}{0.48\textwidth}
    \centering
    \includegraphics[width=\textwidth]{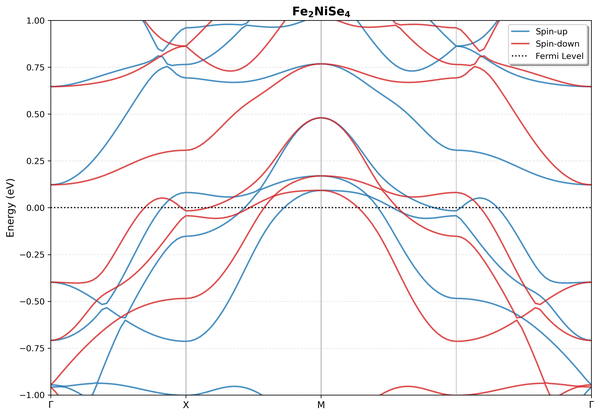}
    \caption*{\small Fe$_2$NiSe$_4$ }
\end{minipage}
\\[1em]
\begin{minipage}{0.48\textwidth}
    \centering
    \includegraphics[width=\textwidth]{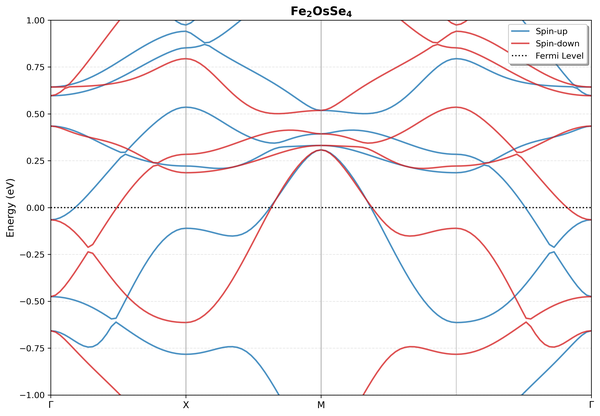}
    \caption*{\small Fe$_2$OsSe$_4$ }
\end{minipage}
\hfill
\begin{minipage}{0.48\textwidth}
    \centering
    \includegraphics[width=\textwidth]{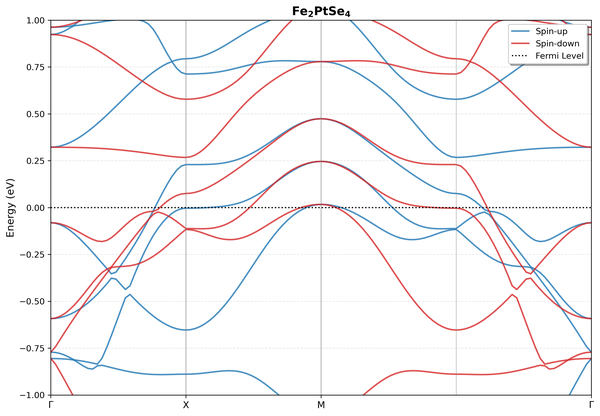}
    \caption*{\small Fe$_2$PtSe$_4$ }
\end{minipage}
\end{figure}

\begin{figure}[H]
\centering
\begin{minipage}{0.48\textwidth}
    \centering
    \includegraphics[width=\textwidth]{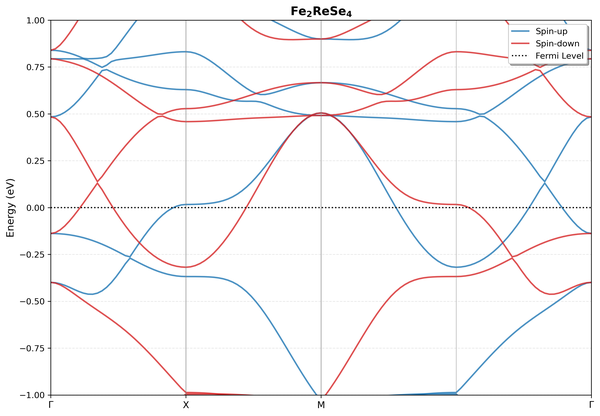}
    \caption*{\small Fe$_2$ReSe$_4$ }
\end{minipage}
\hfill
\begin{minipage}{0.48\textwidth}
    \centering
    \includegraphics[width=\textwidth]{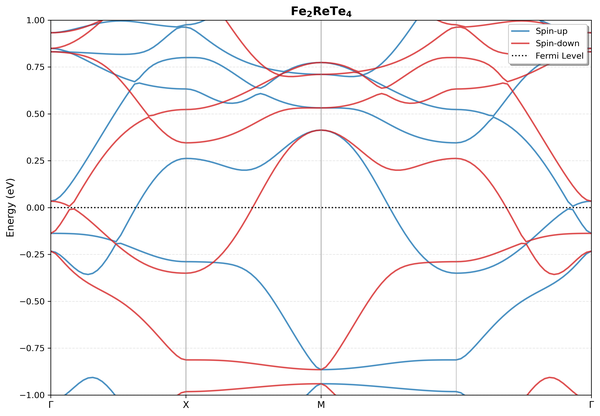}
    \caption*{\small Fe$_2$ReTe$_4$ }
\end{minipage}
\\[1em]
\begin{minipage}{0.48\textwidth}
    \centering
    \includegraphics[width=\textwidth]{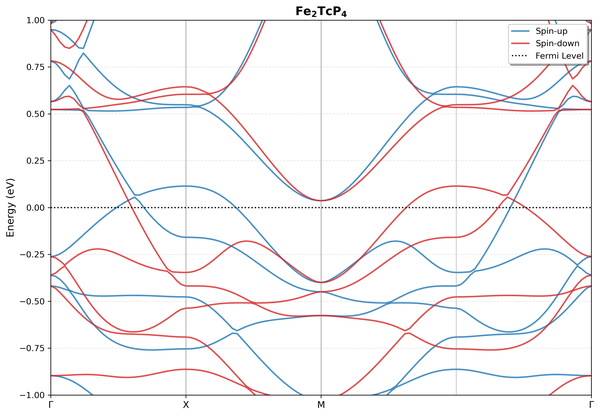}
    \caption*{\small Fe$_2$TcP$_4$ }
\end{minipage}
\hfill
\begin{minipage}{0.48\textwidth}
    \centering
    \includegraphics[width=\textwidth]{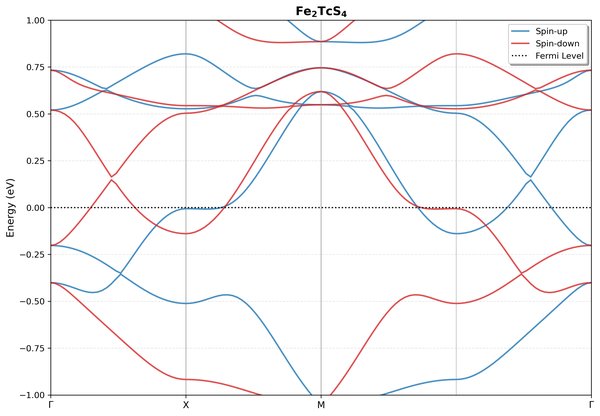}
    \caption*{\small Fe$_2$TcS$_4$ }
\end{minipage}
\\[1em]
\begin{minipage}{0.48\textwidth}
    \centering
    \includegraphics[width=\textwidth]{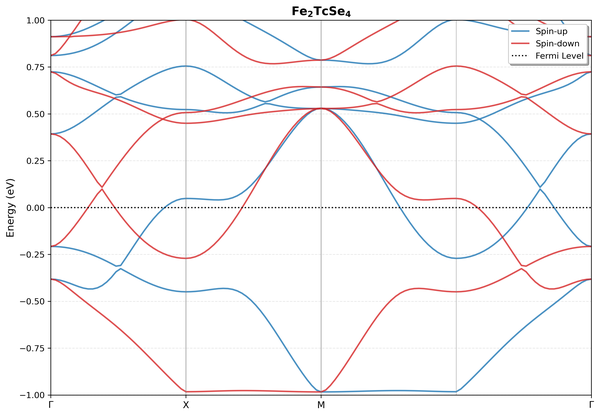}
    \caption*{\small Fe$_2$TcSe$_4$ }
\end{minipage}
\hfill
\begin{minipage}{0.48\textwidth}
    \centering
    \includegraphics[width=\textwidth]{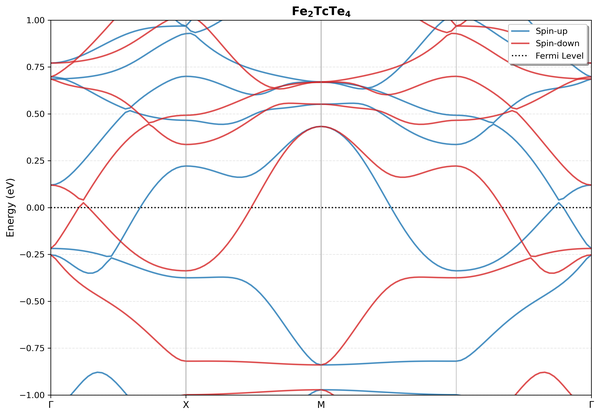}
    \caption*{\small Fe$_2$TcTe$_4$ }
\end{minipage}
\end{figure}

\begin{figure}[H]
\centering
\begin{minipage}{0.48\textwidth}
    \centering
    \includegraphics[width=\textwidth]{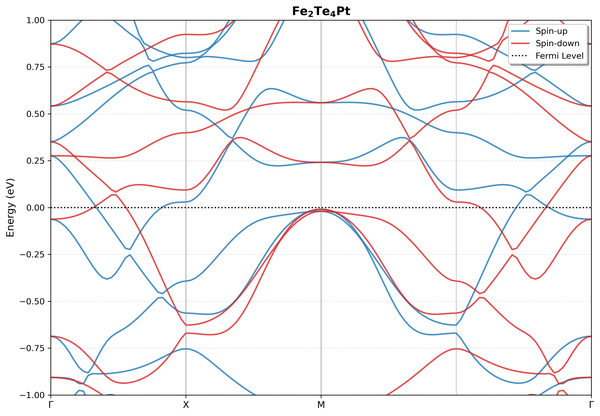}
    \caption*{\small Fe$_2$Te$_4$Pt }
\end{minipage}
\hfill
\begin{minipage}{0.48\textwidth}
    \centering
    \includegraphics[width=\textwidth]{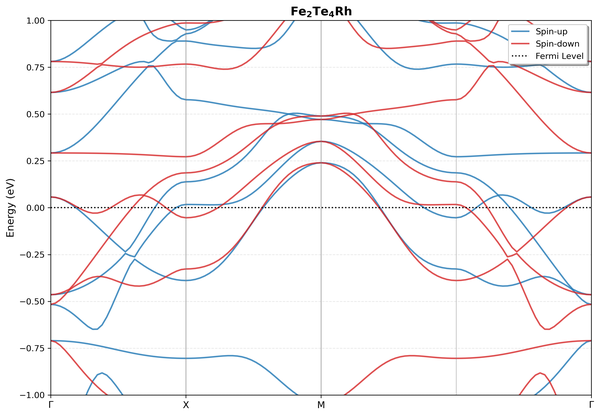}
    \caption*{\small Fe$_2$Te$_4$Rh }
\end{minipage}
\\[1em]
\begin{minipage}{0.48\textwidth}
    \centering
    \includegraphics[width=\textwidth]{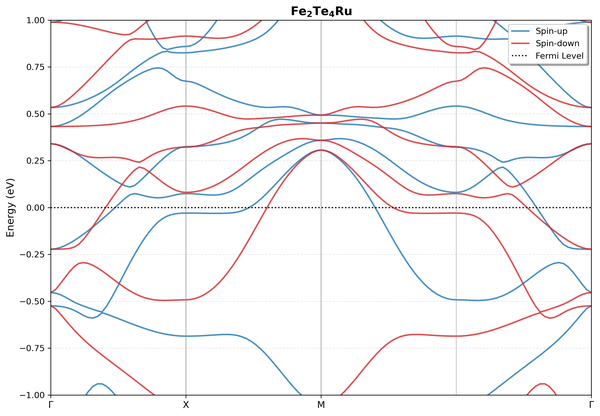}
    \caption*{\small Fe$_2$Te$_4$Ru }
\end{minipage}
\hfill
\begin{minipage}{0.48\textwidth}
    \centering
    \includegraphics[width=\textwidth]{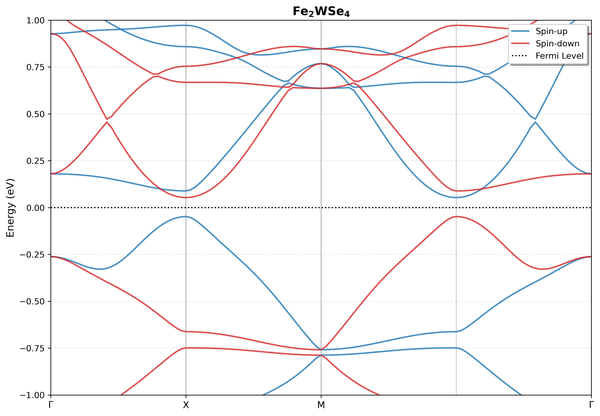}
    \caption*{\small Fe$_2$WSe$_4$ }
\end{minipage}
\\[1em]
\begin{minipage}{0.48\textwidth}
    \centering
    \includegraphics[width=\textwidth]{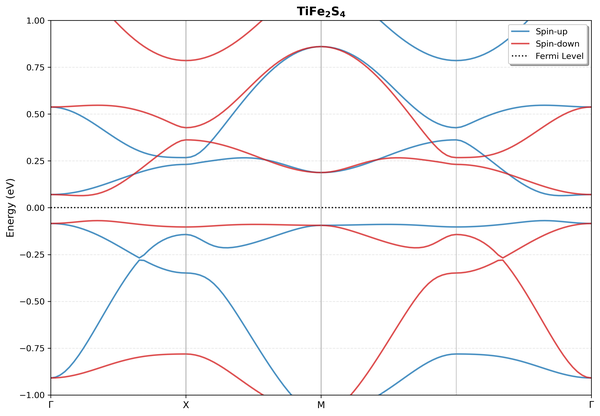}
    \caption*{\small TiFe$_2$S$_4$ }
\end{minipage}
\hfill
\begin{minipage}{0.48\textwidth}
    \centering
    \includegraphics[width=\textwidth]{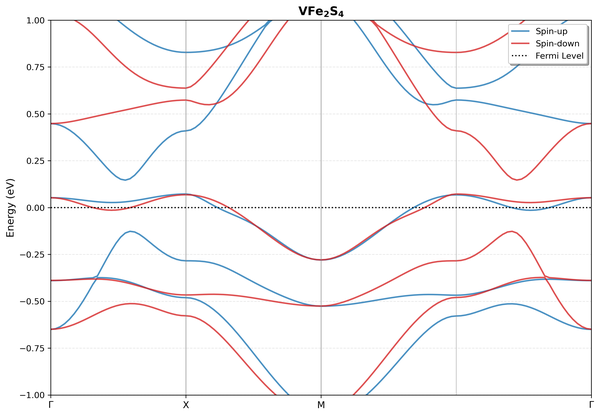}
    \caption*{\small VFe$_2$S$_4$ }
\end{minipage}
\end{figure}

\clearpage


\begin{figure}[H]
\centering
\begin{minipage}{0.48\textwidth}
    \centering
    \includegraphics[width=\textwidth]{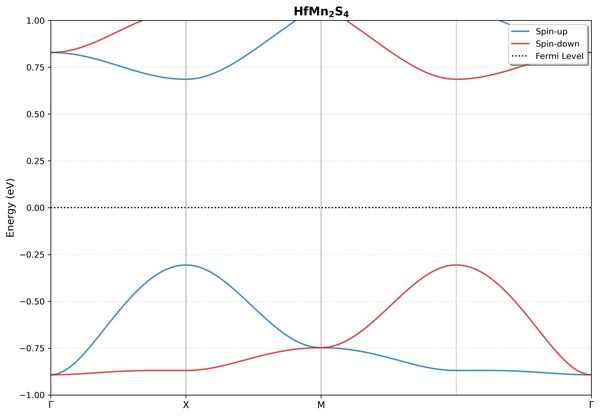}
    \caption*{\small HfMn$_2$S$_4$ }
\end{minipage}
\hfill
\begin{minipage}{0.48\textwidth}
    \centering
    \includegraphics[width=\textwidth]{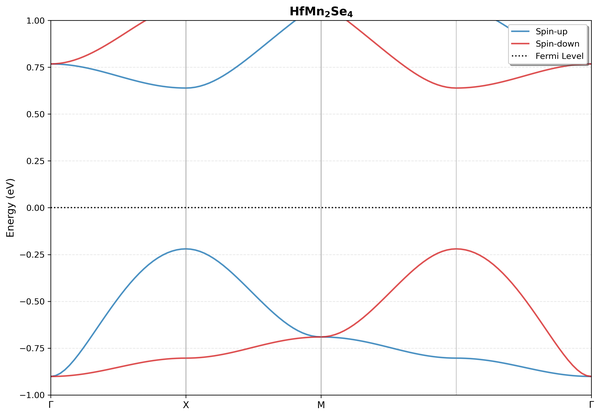}
    \caption*{\small HfMn$_2$Se$_4$ }
\end{minipage}
\\[1em]
\begin{minipage}{0.48\textwidth}
    \centering
    \includegraphics[width=\textwidth]{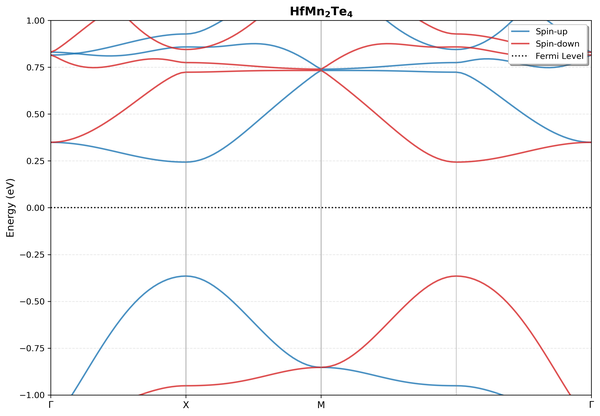}
    \caption*{\small HfMn$_2$Te$_4$ }
\end{minipage}
\hfill
\begin{minipage}{0.48\textwidth}
    \centering
    \includegraphics[width=\textwidth]{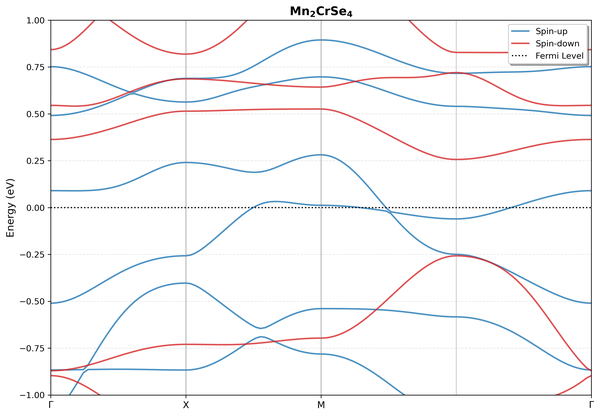}
    \caption*{\small Mn$_2$CrSe$_4$ }
\end{minipage}
\\[1em]
\begin{minipage}{0.48\textwidth}
    \centering
    \includegraphics[width=\textwidth]{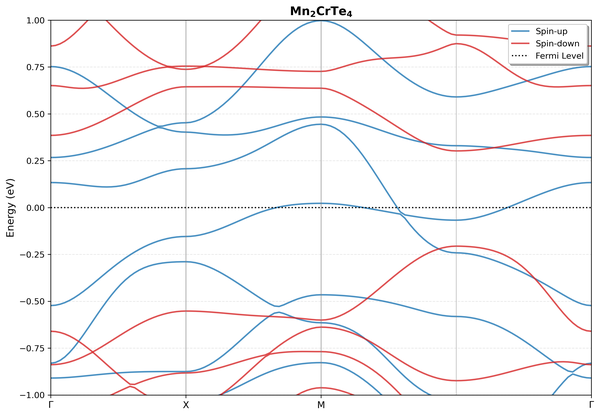}
    \caption*{\small Mn$_2$CrTe$_4$ }
\end{minipage}
\hfill
\begin{minipage}{0.48\textwidth}
    \centering
    \includegraphics[width=\textwidth]{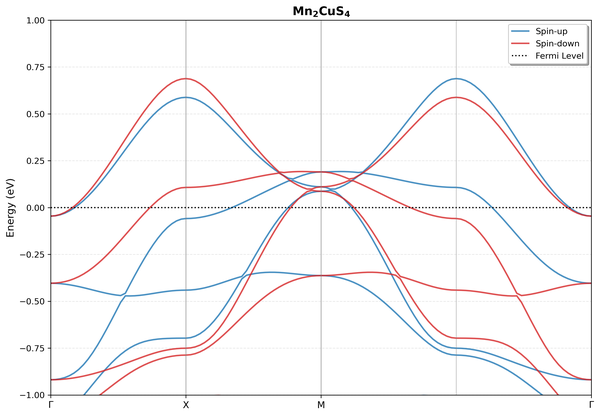}
    \caption*{\small Mn$_2$CuS$_4$ }
\end{minipage}
\end{figure}

\begin{figure}[H]
\centering
\begin{minipage}{0.48\textwidth}
    \centering
    \includegraphics[width=\textwidth]{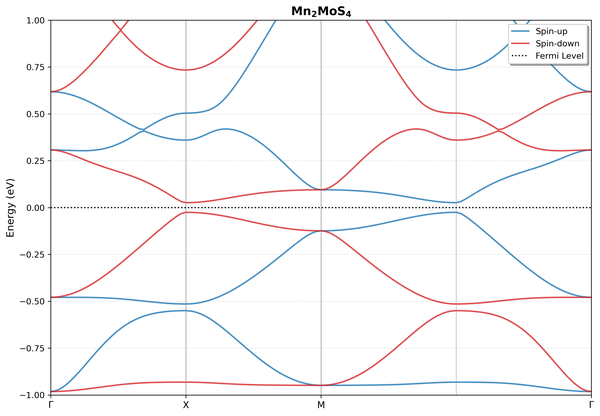}
    \caption*{\small Mn$_2$MoS$_4$ }
\end{minipage}
\hfill
\begin{minipage}{0.48\textwidth}
    \centering
    \includegraphics[width=\textwidth]{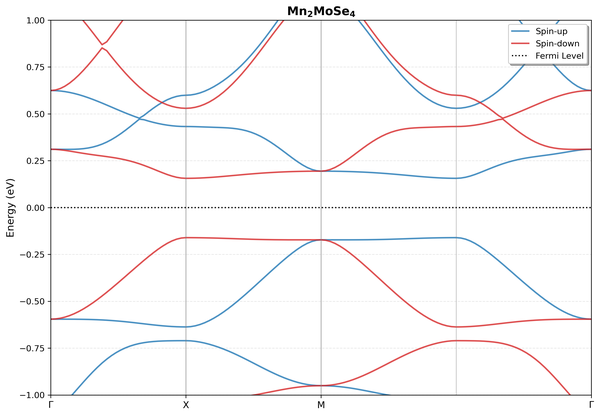}
    \caption*{\small Mn$_2$MoSe$_4$ }
\end{minipage}
\\[1em]
\begin{minipage}{0.48\textwidth}
    \centering
    \includegraphics[width=\textwidth]{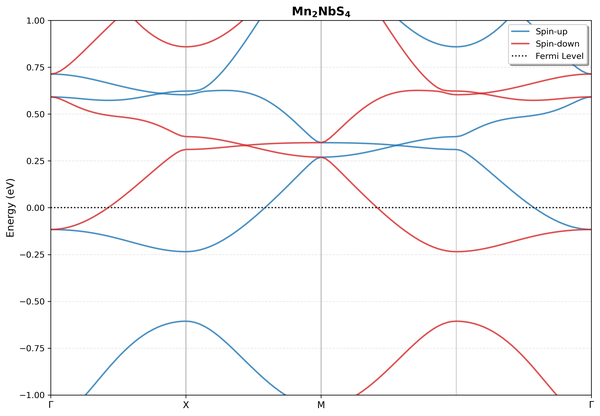}
    \caption*{\small Mn$_2$NbS$_4$ }
\end{minipage}
\hfill
\begin{minipage}{0.48\textwidth}
    \centering
    \includegraphics[width=\textwidth]{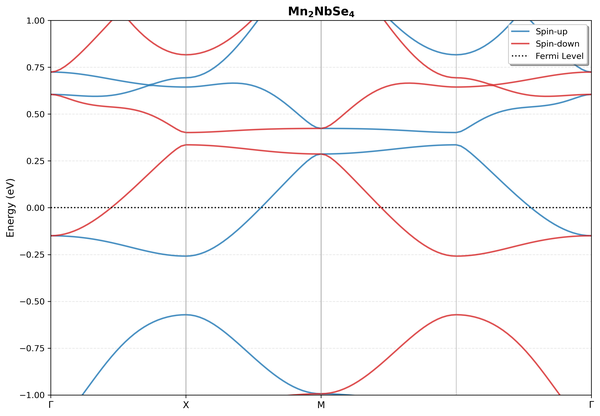}
    \caption*{\small Mn$_2$NbSe$_4$ }
\end{minipage}
\\[1em]
\begin{minipage}{0.48\textwidth}
    \centering
    \includegraphics[width=\textwidth]{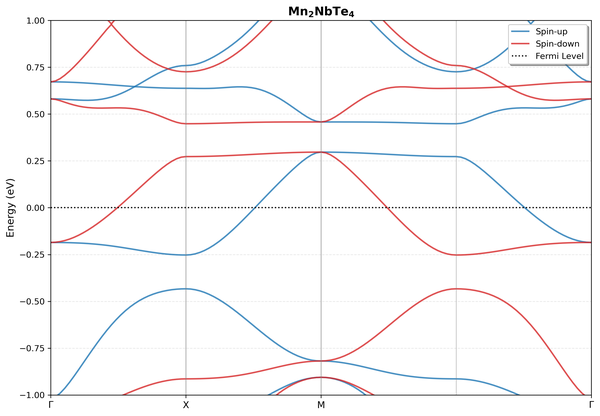}
    \caption*{\small Mn$_2$NbTe$_4$ }
\end{minipage}
\hfill
\begin{minipage}{0.48\textwidth}
    \centering
    \includegraphics[width=\textwidth]{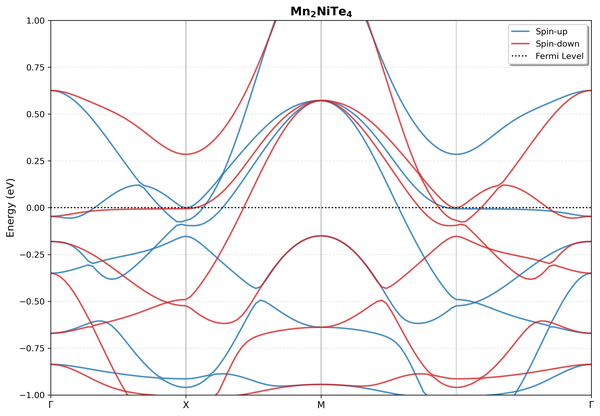}
    \caption*{\small Mn$_2$NiTe$_4$ }
\end{minipage}
\end{figure}

\begin{figure}[H]
\centering
\begin{minipage}{0.48\textwidth}
    \centering
    \includegraphics[width=\textwidth]{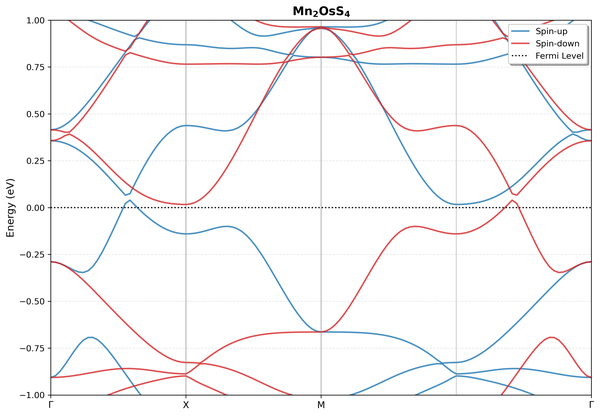}
    \caption*{\small Mn$_2$OsS$_4$ }
\end{minipage}
\hfill
\begin{minipage}{0.48\textwidth}
    \centering
    \includegraphics[width=\textwidth]{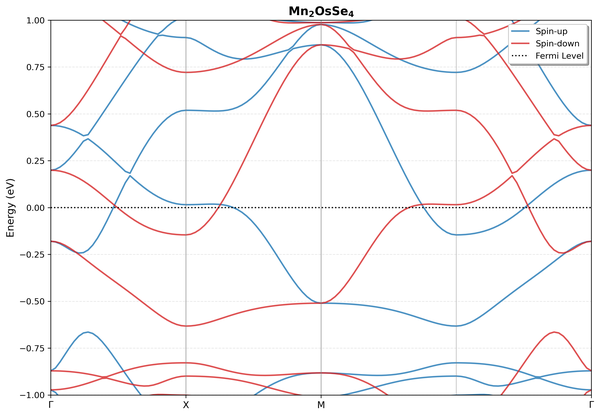}
    \caption*{\small Mn$_2$OsSe$_4$ }
\end{minipage}
\\[1em]
\begin{minipage}{0.48\textwidth}
    \centering
    \includegraphics[width=\textwidth]{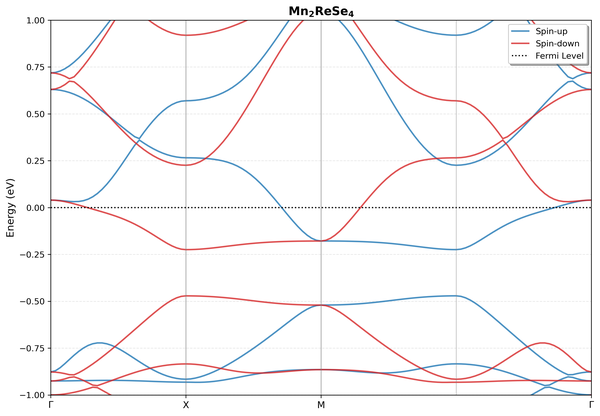}
    \caption*{\small Mn$_2$ReSe$_4$ }
\end{minipage}
\hfill
\begin{minipage}{0.48\textwidth}
    \centering
    \includegraphics[width=\textwidth]{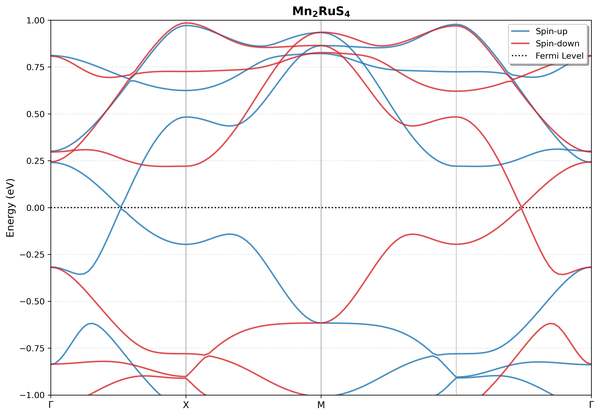}
    \caption*{\small Mn$_2$RuS$_4$ }
\end{minipage}
\\[1em]
\begin{minipage}{0.48\textwidth}
    \centering
    \includegraphics[width=\textwidth]{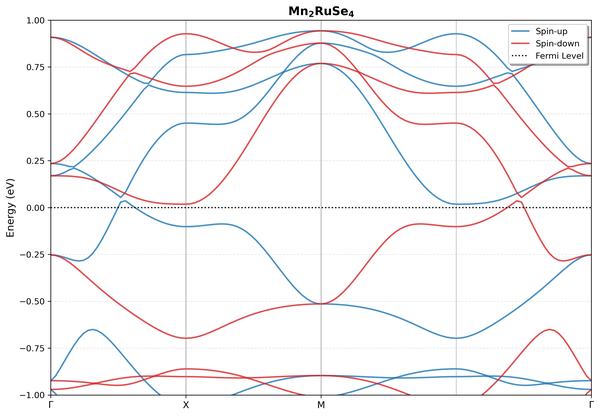}
    \caption*{\small Mn$_2$RuSe$_4$ }
\end{minipage}
\hfill
\begin{minipage}{0.48\textwidth}
    \centering
    \includegraphics[width=\textwidth]{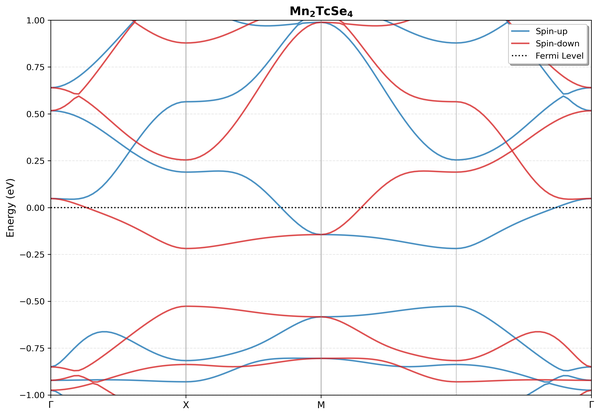}
    \caption*{\small Mn$_2$TcSe$_4$ }
\end{minipage}
\end{figure}

\begin{figure}[H]
\centering
\begin{minipage}{0.48\textwidth}
    \centering
    \includegraphics[width=\textwidth]{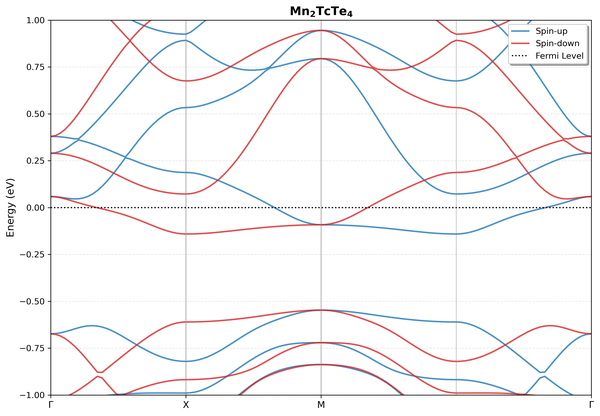}
    \caption*{\small Mn$_2$TcTe$_4$ }
\end{minipage}
\hfill
\begin{minipage}{0.48\textwidth}
    \centering
    \includegraphics[width=\textwidth]{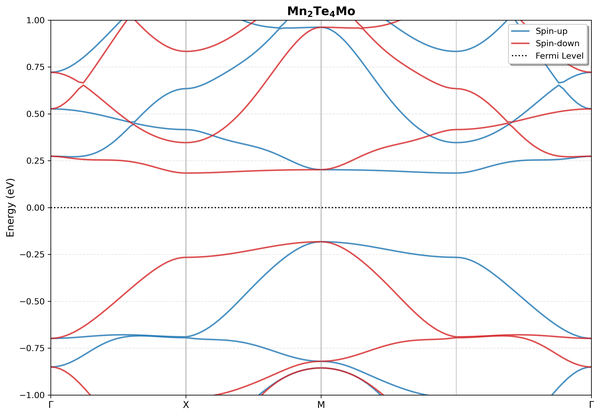}
    \caption*{\small Mn$_2$Te$_4$Mo }
\end{minipage}
\\[1em]
\begin{minipage}{0.48\textwidth}
    \centering
    \includegraphics[width=\textwidth]{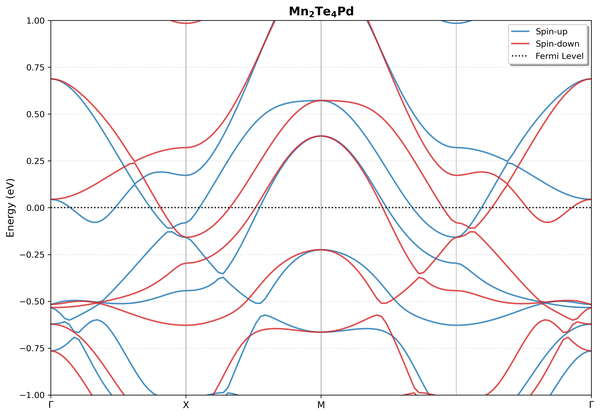}
    \caption*{\small Mn$_2$Te$_4$Pd }
\end{minipage}
\hfill
\begin{minipage}{0.48\textwidth}
    \centering
    \includegraphics[width=\textwidth]{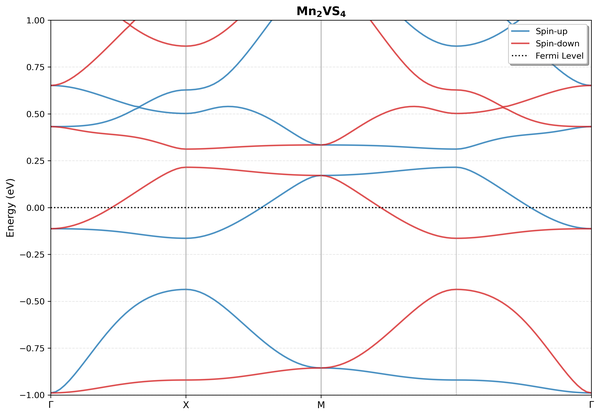}
    \caption*{\small Mn$_2$VS$_4$ }
\end{minipage}
\\[1em]
\begin{minipage}{0.48\textwidth}
    \centering
    \includegraphics[width=\textwidth]{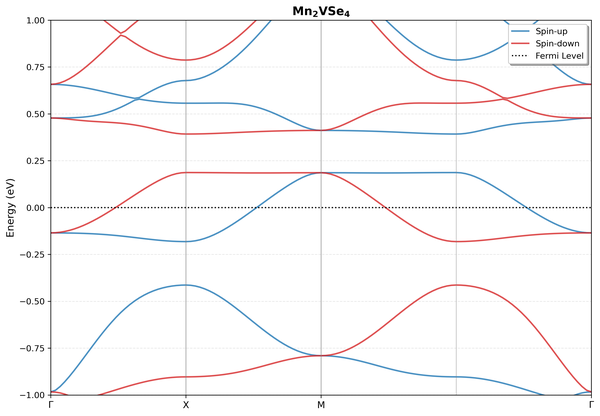}
    \caption*{\small Mn$_2$VSe$_4$ }
\end{minipage}
\hfill
\begin{minipage}{0.48\textwidth}
    \centering
    \includegraphics[width=\textwidth]{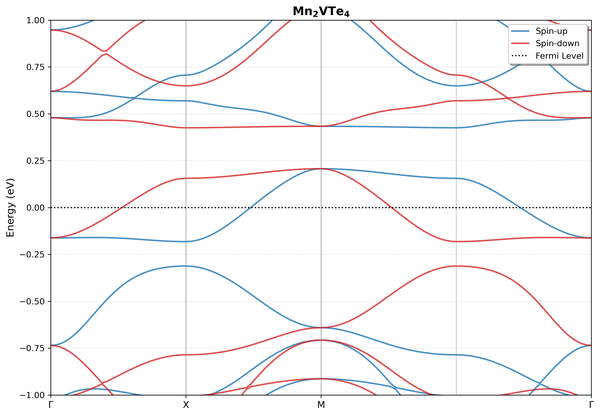}
    \caption*{\small Mn$_2$VTe$_4$ }
\end{minipage}
\end{figure}

\begin{figure}[H]
\centering
\begin{minipage}{0.48\textwidth}
    \centering
    \includegraphics[width=\textwidth]{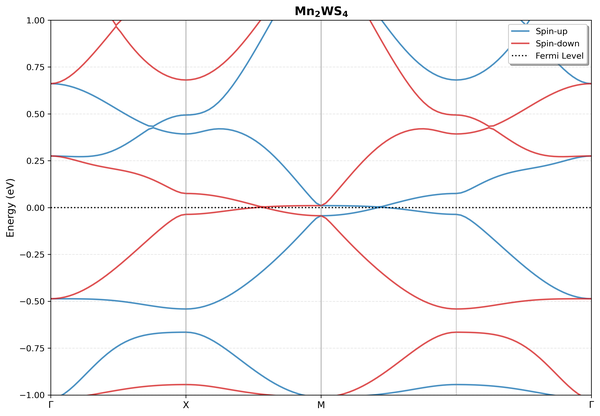}
    \caption*{\small Mn$_2$WS$_4$ }
\end{minipage}
\hfill
\begin{minipage}{0.48\textwidth}
    \centering
    \includegraphics[width=\textwidth]{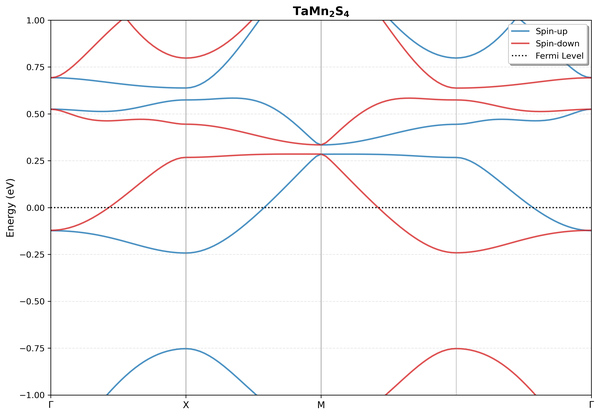}
    \caption*{\small TaMn$_2$S$_4$ }
\end{minipage}
\\[1em]
\begin{minipage}{0.48\textwidth}
    \centering
    \includegraphics[width=\textwidth]{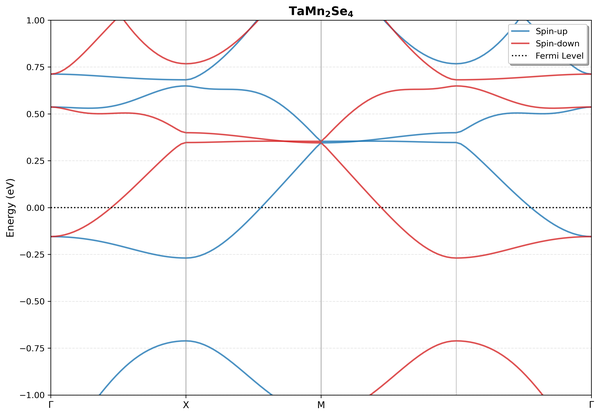}
    \caption*{\small TaMn$_2$Se$_4$ }
\end{minipage}
\hfill
\begin{minipage}{0.48\textwidth}
    \centering
    \includegraphics[width=\textwidth]{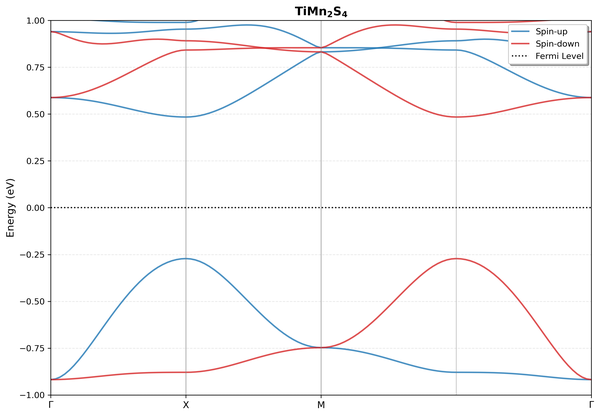}
    \caption*{\small TiMn$_2$S$_4$ }
\end{minipage}
\\[1em]
\begin{minipage}{0.48\textwidth}
    \centering
    \includegraphics[width=\textwidth]{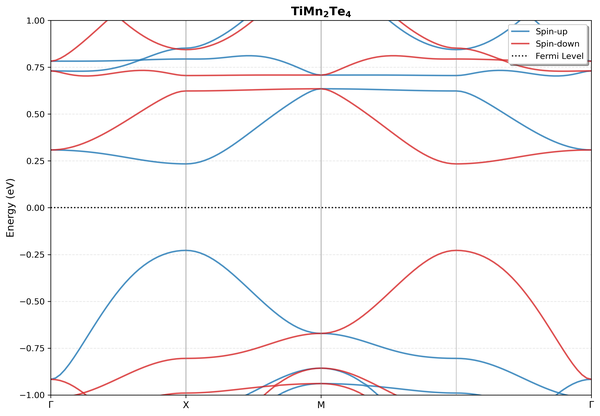}
    \caption*{\small TiMn$_2$Te$_4$ }
\end{minipage}
\end{figure}

\clearpage


\begin{figure}[H]
\centering
\begin{minipage}{0.48\textwidth}
    \centering
    \includegraphics[width=\textwidth]{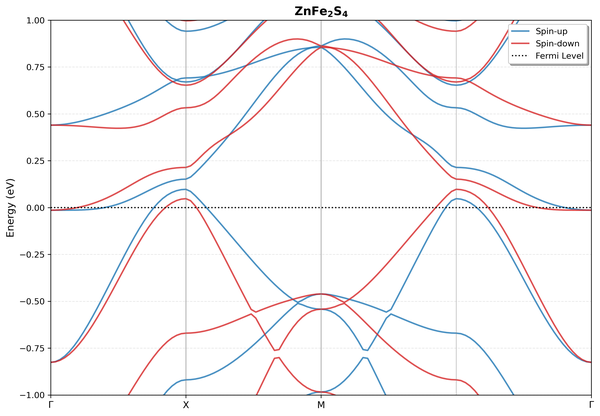}
    \caption*{\small ZnFe$_2$S$_4$ }
\end{minipage}
\hfill
\begin{minipage}{0.48\textwidth}
    \centering
    \includegraphics[width=\textwidth]{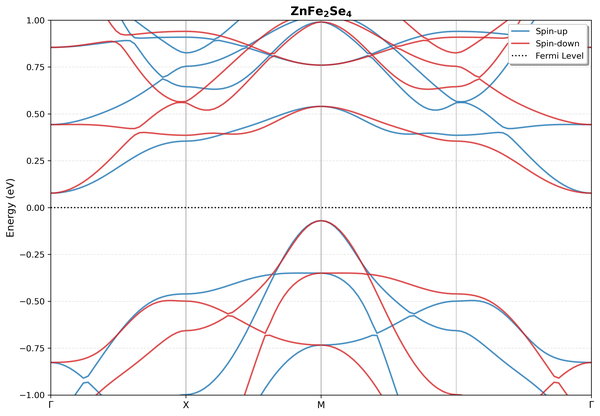}
    \caption*{\small ZnFe$_2$Se$_4$ }
\end{minipage}
\\[1em]
\begin{minipage}{0.48\textwidth}
    \centering
    \includegraphics[width=\textwidth]{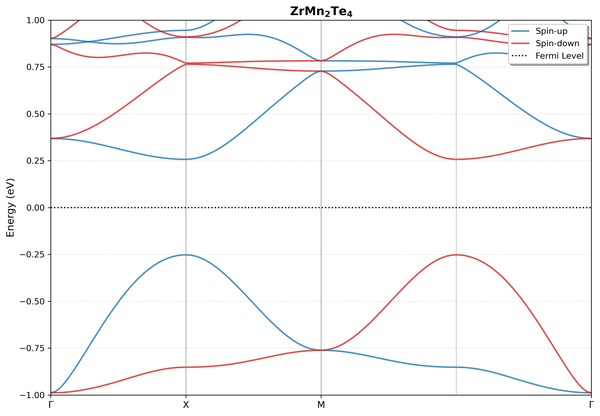}
    \caption*{\small ZrMn$_2$Te$_4$ }
\end{minipage}
\end{figure}

\clearpage

\section{Machine learning interatomic potential models supported by UniMatSim}

\begin{table}[htbp]
\centering
\caption{Machine learning interatomic potential models supported by UniMatSim}
\label{tab:potential-models}
\begin{tabularx}{\textwidth}{@{}llX@{}}
\toprule
\textbf{Model name} & \textbf{Identifier} & \textbf{GitHub} \\
\midrule
CHGNet & PotentialModelType.CHGNET & \href{https://github.com/CederGroupHub/chgnet}{CederGroupHub/chgnet} \\
M3GNet & PotentialModelType.M3GNET & \href{https://github.com/materialsvirtuallab/m3gnet}{materialsvirtuallab/m3gnet}\\
MACE & PotentialModelType.MACE & \href{https://github.com/ACEsuit/mace}{ACEsuit/mace}\\
MatterSim & PotentialModelType.MATTERSIM & \href{https://github.com/microsoft/mattersim}{microsoft/mattersim}\\
SevenNet & PotentialModelType.SEVENNET & \href{https://github.com/MDIL-SNU/SevenNet}{MDIL-SNU/SevenNet} \\
\bottomrule
\end{tabularx}
\end{table}

\end{document}